\documentstyle[preprint,aps]{revtex}
\tighten
\draft
\widetext
\input epsf
\preprint{YITP-01-03}
\bigskip
\bigskip

\begin{document}
\title{Brane-Bulk Duality and Non-conformal Gauge Theories}
\medskip
\author{Zurab Kakushadze\footnote{E-mail: 
zurab@insti.physics.sunysb.edu} and Radu Roiban\footnote{E-mail:
roiban@insti.physics.sunysb.edu}}
\bigskip
\address{C.N. Yang Institute for Theoretical Physics\\ 
State University of New York, Stony Brook, NY 11794}

\date{February 19, 2001}
\bigskip
\medskip
\maketitle

\begin{abstract} 
{}We discuss non-conformal gauge theories from type IIB D3-branes embedded
in orbifolded space-times. Such theories can be obtained by allowing some
non-vanishing logarithmic twisted tadpoles. 
In certain cases with ${\cal N}=0,1$
supersymmetry correlation functions in the planar limit are the same as in the
parent ${\cal N}=2$ supersymmetric theories. In particular, the effective 
action in such theories perturbatively is not renormalized
beyond one loop in the planar limit. 
In the ${\cal N}=2$ as well as such ${\cal N}=0,1$ theories
quantum corrections in the D3-brane gauge theories are encoded in the 
corresponding
classical higher dimensional field theories whose actions contain the 
twisted fields with non-vanishing tadpoles. We argue that this duality can be
extended to the non-perturbative level in the ${\cal N}=2$ theories. We
give some evidence that this might also be the case for ${\cal N}=0,1$ 
theories as well.  
\end{abstract}

\section{Introduction}

{}In 't Hooft's large $N$ limit \cite{tHooft} gauge theories are expected to be
drastically simplified. Thus, in this limit the gauge theory diagrams are
organized in terms of Riemann surfaces, where each extra handle on the 
surface suppresses the corresponding diagram by 
$1/N^2$. The large $N$ expansion,
therefore, resembles perturbative expansion in string theory. In the case
of four-dimensional gauge theories this connection can be made precise in
the context of type IIB string theory in the presence of a large number $N$
of D3-branes \cite{z1}\footnote{The generalization of this setup in the
presence of orientifold planes was discussed in \cite{z2}.}. 
Thus, we consider a limit where 
$\alpha'\rightarrow 0$, $g_s\rightarrow 0$ and $N\rightarrow \infty$, 
while keeping $\lambda \equiv N g_s$ fixed, where $g_s$ is the type IIB
string coupling. Note that in this context a world-sheet with $g$
handles and $b$ boundaries is weighted with
\begin{equation}
 (N g_s)^b g_s^{2g-2}=\lambda^{2g-2+b} N^{-2g+2}~.
\end{equation}
Once we identify $g_s=g_{\rm{\small YM}}^2$, this is the same as the large 
$N$ expansion considered by 't Hooft. 
Note that for this expansion to make sense we must
keep $\lambda$ at a small value $\lambda <1$.
In this regime we can map the string diagrams directly to
(various sums of) large $N$ Feynman diagrams.
Note, in particular, that the genus $g=0$ planar diagrams
dominate in the large $N$ limit\footnote{Note that if $\lambda >1$, then
no matter how large $N$ is, for sufficiently many boundaries the
higher genus terms become relevant, and we lose the genus expansion. In fact,
in this regime one expects
an effective supergravity description to take over as discussed in
\cite{malda,GKP,Witten}.}.

{}If the space transverse to the D3-branes in the setup of \cite{z1} is 
${\bf R}^6$, then we obtain the ${\cal N}=4$ supersymmetric $U(N)$ gauge theory
on the D3-branes, which is conformal. On the other hand, we can also consider
orbifolds of ${\bf R}^6$, which leads to gauge theories with reduced 
supersymmetry\footnote{The $\lambda>1$ versions of these orbifold theories
via the compactifications of type IIB on AdS$_5\times ({\bf S}^5/\Gamma)$
(where $\Gamma$ is the orbifold group) were originally discussed in 
\cite{Kachru}.}. As was shown in \cite{z1}, if we cancel all twisted tadpoles
in such models, in the large $N$ limit the corresponding ${\cal N}=0,1,2$
gauge theories are conformal. Moreover, in the planar limit 
the (on-shell) correlation functions
in such theories are the same as in the parent ${\cal N}=4$ gauge theory.   

{}In this paper we discuss non-conformal gauge theories within the setup of
\cite{z1}. Such theories can be 
obtained by allowing some twisted tadpoles to be 
non-vanishing. In particular, we can have consistent embeddings
of non-conformal gauge theories if we allow logarithmic tadpoles, 
which correspond to the twisted sectors with fixed point loci of real dimension
two. In particular, even though the corresponding string backgrounds are
not finite (in the sense that we have logarithmic ultra-violet divergences),
they are still consistent as far as the gauge theories are concerned, 
and the divergences correspond to the running in the
four-dimensional gauge theories on the D3-branes.

{}Regularization of the aforementioned divergences can be conveniently 
discussed in the context of what we refer to as the brane-bulk duality,
which is a consequence of the open-closed string duality. 
In particular, in certain non-trivial ${\cal N}=0,1$ cases in the planar limit
the corresponding gauge theories perturbatively are not renormalized beyond 
one-loop. In fact, in this limit the (on-shell) correlation functions in these
theories are the same as in the parent ${\cal N}=2$ non-conformal gauge 
theories. In the ${\cal N}=2$ as well as the aforementioned ${\cal N}=0,1$
cases the brane-bulk duality is particularly simple, and implies that the
quantum corrections in the corresponding gauge theories are encoded in 
classical higher dimensional field theories whose actions contain the 
twisted fields with non-vanishing tadpoles. In particular, various quantum 
corrections can be obtained via integrating out the bulk fields in the
corresponding classical action, 
that is, by considering the self-interaction of the D3-branes
via the bulk fields. We give explicit computations in various 
${\cal N}=0,1,2$ examples in this context, including the treatment of 
divergences.

{}We also discuss whether the brane-bulk duality can be extended to the 
non-perturbative level in the aforementioned theories. In the ${\cal N}=2$ 
cases we argue that, since we are working in the large $N$ limit, the low 
energy effective action does not receive non-perturbative corrections. We also
conjecture that this should be the case for the corresponding ${\cal N}=0,1$ 
theories as well. In the ${\cal N}=1$ cases we verify that there are no
non-perturbative corrections to the superpotential in these theories in the
large $N$ limit.

{}The remainder of this paper is organized as follows. In section II we discuss
our setup. In section III we discuss non-conformal large $N$ gauge theories
which can be constructed within this setup. In section IV we discuss the
large $N$ limit and brane-bulk-duality. In sections V, VI and VII we give
details of classical computations that in the context of the brane-bulk duality
reproduce quantum results in the corresponding ${\cal N}=2$, ${\cal N}=1$
and ${\cal N}=0$ gauge theories, respectively. In section VIII 
we comment on the non-perturbative extension of the brane-bulk duality. 
In section IX we give a few concluding remarks. In Appendix A we compute
the brane-bulk couplings used in sections V, VI and VII. 

\section{Setup}

{}In this section we discuss the setup within which we will consider 
four-dimensional large $N$ gauge theories in the context of brane-bulk duality.
Parts of our discussion in this section closely follow \cite{z1}. 
Thus, consider type IIB string theory in the presence of $N$ coincident  
D3-branes with the space transverse to the D-branes 
${\cal M}={\bf R}^6/\Gamma$. The orbifold group 
$\Gamma=\{g_a|a=1,\dots,|\Gamma|\}$ ($g_1=1$) must be 
a finite discrete subgroup of $Spin(6)$. If $\Gamma\subset SU(3)(SU(2))$, 
we have ${\cal N}=1$ (${\cal N}=2$) unbroken supersymmetry, and ${\cal N}=0$,
otherwise.

{}We will confine our attention to the cases where type IIB on ${\cal M}$ is
a modular invariant theory\footnote{This is always the case if there are
some unbroken supersymmetries. If all supersymmetries are broken, this is also
true if $\not\exists{\bf Z}_2\subset\Gamma$. If 
$\exists{\bf Z}_2\subset\Gamma$, then modular invariance requires 
that the set of points in ${\bf R}^6$
fixed under the ${\bf Z}_2$ twist has real dimension 2.}. The action of the
orbifold on
the coordinates $X_i$ ($i=1,\dots,6$) on ${\cal M}$ can be described
in terms of $SO(6)$ matrices:
$g_a:X_i\rightarrow (g_a)_{ij} X_j$.
The world-sheet fermionic superpartners of $X_i$
transform in the same way.
We also need to specify
the action of the orbifold group on the Chan-Paton charges carried by the
D3-branes. It is described by $N\times N$ matrices $\gamma_a$ that
form a representation of $\Gamma$. Note that $\gamma_1$ is an
identity matrix, and ${\rm Tr}(\gamma_1)=N$.

{}The D-brane sector of the theory is described by an oriented open
string theory. In particular, the world-sheet expansion corresponds
to summing over oriented Riemann surfaces with arbitrary genus $g$ and
arbitrary
number of boundaries $b$, where the boundaries of the world-sheet are mapped
to the D3-brane world-volume. Moreover, we must consider various ``twists''
around the cycles of the Riemann surface. The choice of these ``twists''
corresponds to a choice of homomorphism of the fundamental group of
the Riemann surface with boundaries to $\Gamma$.

{}For example, consider the one-loop vacuum amplitude ($g=0, b=2$). The
corresponding graph is an annulus whose boundaries lie on D3-branes.
The annulus amplitude is given by
\begin{equation}
 {\cal C}=\int_0^\infty {dt\over t} ~{\cal Z}~.
\end{equation}
The one-loop partition function 
${\cal Z}$ in the light-cone gauge is given by
\begin{equation}\label{partition}
 {\cal Z}={1\over |\Gamma|}\sum_a
 {\rm Tr}  \left( g_a ~{{1-(-1)^F}\over 2}~
 e^{-2\pi tL_0}
 \right)~,
\end{equation}
where $L_0$ is the light-cone Hamiltonian,
$F$ is the fermion number operator, $t$ is the real modular parameter
on the cylinder, and the trace includes sum over the Chan-Paton factors.
The states in the Neveu-Schwarz (NS) sector are space-time bosons and enter the
partition function with weight $+1$, whereas the states in the Ramond (R)
sector are space-time fermions and contribute with weight $-1$.

{}The elements $g_a$ acting in the Hilbert space of open strings
act both on the left end and the right end of the open string.
This action
corresponds to $\gamma_a\otimes \gamma_a$ acting on the Chan-Paton indices.
The partition function (\ref{partition}), therefore,
has the following form:
\begin{equation}\label{gsq}
 {\cal Z}={1\over |\Gamma|}
 \sum_a \left({\rm Tr}(\gamma_a)\right)^2 {\cal Z}_a~,
\end{equation}
where ${\cal Z}_a$ are characters
corresponding to the world-sheet degrees of freedom. The ``untwisted''
character
${\cal Z}_1$ is the same as in the ${\cal N}=4$ theory for which
$\Gamma=\{1\}$. The information about the fact that the orbifold theory
has reduced supersymmetry is encoded in the ``twisted'' characters
${\cal Z}_a$, $a\not=1$.

\subsection{Tadpole Cancellation}

{}In this subsection we discuss one-loop tadpoles arising in the above 
setup. As was pointed out in \cite{z1}, if all tadpoles are canceled, then
the resulting theory is finite in the large $N$ limit. However, not all 
tadpoles need to be canceled to have a consistent four-dimensional gauge 
theory. In fact, we can obtain non-conformal gauge theories if we allow such
tadpoles.

{}The characters ${\cal Z}_a$ in (\ref{gsq}) are given by
\begin{equation}
 {\cal Z}_a={1\over (8\pi^2\alpha^\prime t)^2 }{1\over 
 \left[\eta(e^{-2\pi t})\right]^{2+d_a}}
 \left[{\cal X}_a(e^{-2\pi t})-{\cal Y}_a(e^{-2\pi t})\right]~,
\end{equation}
where $d_a$ is the real dimension of the set of
points in ${\bf R}^6$ fixed under the twist $g_a$. The factor of $(8\pi^2
\alpha^\prime t)^2$ in the denominator comes from the bosonic zero modes 
corresponding to four directions along the D3-brane 
world-volume. Two of the $\eta$-functions come from the
bosonic
oscillators corresponding to two spatial directions along the D3-brane 
world-volume (the time-like and longitudinal contributions are
absent as we are working in the light-cone gauge). The other $d_a$
$\eta$-functions come from the bosonic oscillators corresponding
to the directions transverse to the D-branes untouched by the orbifold
twist $g_a$. Finally, the
characters ${\cal X}_a$, ${\cal Y}_a$ correspond to the contributions of
the world-sheet fermions, as well as the
world-sheet bosons with $g_a$ acting non-trivially
on them (for $a\not=1$): 
\begin{eqnarray}
 &&{\cal X}_a ={1\over 2}{\rm Tr}^\prime \left[g_a e^{-2\pi t L_0}\right]~,\\
 &&{\cal Y}_a ={1\over 2}
 {\rm Tr}^\prime \left[g_a (-1)^F e^{-2\pi t L_0}\right]~,
\end{eqnarray}
where prime in ${\rm Tr}^\prime$ indicates that the trace is restricted as
described above.

{}For the annulus amplitude we therefore have
\begin{equation}
 {\cal C}={1\over (8\pi^2\alpha^\prime)^2 }{1\over |\Gamma|} \sum_a 
 \left[{\cal A}_a - 
 {\cal B}_a\right]~,
\end{equation}
where
\begin{eqnarray}
 &&{\cal A}_a=\left({\rm Tr}\left(\gamma_a\right)\right)^2 
 \int_0^\infty {dt\over t^3}
 {1\over \left[\eta(e^{-2\pi t})\right]^{2+d_a}}{\cal X}_a(e^{-2\pi t})~,\\
 &&{\cal B}_a=\left({\rm Tr}\left(\gamma_a\right)\right)^2 
 \int_0^\infty {dt\over t^3}
 {1\over \left[\eta(e^{-2\pi t})\right]^{2+d_a}}{\cal Y}_a(e^{-2\pi t})~.
\end{eqnarray}
These integrals\footnote{For space-time supersymmetric
theories the total tadpoles vanish: ${\cal A}_a-{\cal B}_a=0$. (The entire
partition function vanishes as the numbers of
space-time bosons and fermions are equal.) For consistency, however, we must
extract tadpoles from individual contributions ${\cal A}_a$ and ${\cal B}_a$.
Thus, for instance, cancellation of certain tadpoles coming from ${\cal B}_a$
is required for consistency of the equations of motion
for the twisted R-R four-form which couples to D3-branes (see below).}
are generically divergent as $t\rightarrow 0$ reflecting the 
presence of tadpoles.
To extract these divergences we can change variables $t=1/\ell$ so that the
divergences correspond to $\ell\rightarrow\infty$:
\begin{eqnarray}\label{NSNS}
 &&{\cal A}_a =\left({\rm Tr}(\gamma_a)\right)^2 \int_0^\infty
 {d\ell\over \ell^{d_a/2}}
  \sum_{\sigma_a} N_{\sigma_a} e^{-2\pi\ell \sigma_a}~,\\
 \label{RR}
 &&{\cal B}_a =\left({\rm Tr}(\gamma_a)\right)^2 \int_0^\infty
  {d\ell\over \ell^{d_a/2}}
 \sum_{\rho_a} N_{\rho_a}e^{-2\pi\ell \rho_a}~.
\end{eqnarray}
The closed string states contributing to ${\cal A}_a$ (${\cal B}_a$)
in the transverse channel
are the NS-NS (R-R) states with $L_0={\overline L}_0=\sigma_a (\rho_a)$
(and $N_{\sigma_a} (N_{\rho_a})>0$ is
the number of such states). The massive states with
$\sigma_a(\rho_a)>0$
do not lead to divergences as $\ell\rightarrow\infty$.
On the other hand, the divergence property of the above integrals
in the $\ell\rightarrow\infty$ limit is determined by the value of $d_a$.
Given the orientability of $\Gamma$ the allowed values of 
$d_a$ are $0,2,4,6$. For $d_1=6$ there is no divergence in ${\cal B}_1$, so
we have no restriction for ${\rm Tr}(\gamma_1)=N$. For $d_a=4$ the
corresponding twisted NS-NS closed string sector contains tachyons. 
This leads to a tachyonic divergence in ${\cal A}_a$ unless 
${\rm Tr}(\gamma_a)=0$ for the corresponding $g_a$ twisted sector.
Finally, the ground states in the R-R sectors are massless, so we 
get divergences due to massless 
R-R states in the integral in ${\cal B}_a$ for large $\ell$
for $d_a=0,2$. In such sectors in non-supersymmetric cases we 
can also have tachyonic
NS-NS divergences, while in supersymmetric cases we have
massless NS-NS divergences.   

{}To avoid complications with tachyons, for now we will focus on
supersymmetric theories (we will discuss non-supersymmetric cases in section
VII). 
We must therefore consider massless tadpoles
arising for $d_a=0,2$. For $d_a=0$ the corresponding integrals are
linearly divergent with $\ell$ as $\ell\rightarrow\infty$. To cancel
such a tadpole we must require that ${\rm Tr}\left(\gamma_a\right)=0$
for the corresponding $g_a$ twisted sector. On the other hand, if such
a tadpole is not canceled, in the four-dimensional field theory language this 
would correspond to having a {\em quadratic} (in the momentum) divergence at 
the one-loop order. This would imply that the corresponding four-dimensional
background is actually inconsistent in the sense that either four-dimensional 
supersymmetry\footnote{Note that if $\Gamma\subset SU(3)$ contains a twist
with $d_a=0$, we have ${\cal N}=1$ supersymmetry.} 
and/or Poincar{\'e}
invariance must be broken\footnote{Since the corresponding twisted
fields $\sigma_a$ propagate in four dimensions, in the presence of the tadpoles
their equations of motion read $\partial^\mu\partial_\mu\sigma_a=c_a+\dots$ 
($\mu=0,1,2,3$, $c_a$ are the tadpoles, and the ellipses stand for terms of
higher power in $\sigma_a$), so that we can {\em a priori} have solutions
with broken four-dimensional Poincar{\'e} invariance.}. 
This is related to the fact that 
in such cases the corresponding twisted closed string states, which
propagate only in the D3-brane world-volume as they are supported at the 
orbifold fixed {\em points} in ${\bf R}^6$, have inconsistent field equations
(yet they couple to the gauge/matter fields describing the low energy
limit of the D3-brane world-volume theory). In fact, as was pointed out in
\cite{z1}, generically uncancelled tadpoles arising in the $d_a=0$ cases 
lead to
non-Abelian gauge {\em anomalies} in the corresponding D3-brane gauge 
theories. Thus, let us consider the following example. 
Let ${\cal M}={\bf C}^3/{\Gamma}$, where
the action of $\Gamma\approx{\bf Z}_3$ 
on the complex coordinates $z_\alpha$ ($\alpha=1,2,3$)
on ${\cal M}$ is that of the Z-orbifold: $g:z_\alpha\rightarrow \omega
z_\alpha$ (where $g$ is the generator of $\Gamma$, and 
$\omega\equiv e^{2\pi i/3}$). Next, let us
choose the representation of $\Gamma$ when acting on the Chan-Paton charges
as follows: $\gamma_g={\rm diag}({\bf I}_{N_1},\omega {\bf I}_{N_2},
\omega^2 {\bf I}_{N_3})$
(where $N_1+N_2+N_3=N$, and ${\bf I}_m$ is an $m\times m$ identity matrix).
The massless spectrum of this model is the ${\cal N}=1$ supersymmetric
$U(N_1)\otimes U(N_2)\otimes U(N_3)$ gauge theory with the matter
consisting of chiral supermultiplets in the following representations:
\begin{equation}
 3\times ({\bf N}_1,{\overline {\bf N}}_2,{\bf 1})~,~~~
     3\times ({\bf 1}, {\bf N}_2,{\overline {\bf N}}_3)~,~~~
     3\times ({\overline {\bf N}}_1,{\bf 1},{\bf N}_3)~.
\end{equation}
Note that this spectrum is anomalous for non-vanishing $N_1,N_2,N_3$
(the non-Abelian gauge anomaly does not cancel)
unless $N_1=N_2=N_3$. On the other hand, ${\rm Tr}(\gamma_g )=0$ if and only if
$N_1=N_2=N_3$. Here we should mention that not all the choices of 
$\gamma_a$ that do not satisfy ${\rm Tr}(\gamma_a )=0$ for $d_a=0$
lead to such apparent inconsistencies. Thus, consider the
same $\Gamma$ as above but with $\gamma_g={\bf I}_N$. The massless spectrum of
this
model is the ${\cal N}=1$ supersymmetric
$U(N)$ gauge theory with no matter, so it is anomaly
free. In fact, for any orbifold group $\Gamma$ containing a twist $g_a$ with
$d_a=0$ we obtain an anomaly free theory only if ${\rm Tr}(\gamma_a)=0$
or $\gamma_a=I_N$. In the latter case, however, as we discussed above,
the corresponding background
is nonetheless inconsistent, so we must require that for such twists  
${\rm Tr}(\gamma_a)=0$.

{}Finally, let us discuss $d_a=2$ cases. For such twists the corresponding
integrals are only logarithmically divergent as $\ell\rightarrow\infty$.
If such a tadpole is not canceled, that is, if the corresponding
${\rm Tr}(\gamma_a)\not=0$, in the four-dimensional field theory language 
this corresponds to having a {\em logarithmic} divergence (in the momentum)
at the one-loop order. As we will see in the following, these logarithmic 
divergences are precisely related to the running in the corresponding gauge
theories, which are {\em not} conformal (even in the large $N$ limit).
Note that the corresponding twisted closed string states now propagate in
two extra dimensions, so that the four-dimensional backgrounds
are perfectly consistent - the tadpoles for these fields simply imply that 
these fields have non-trivial (logarithmic)
profiles in these two extra dimensions, while
the four dimensions along the D-brane world-volume are still flat
(and the four-dimensional supersymmetry is unbroken).   
In fact, the presence of such tadpoles does not introduce any 
anomalies\footnote{This was shown for $\Gamma\approx {\bf Z}_m\otimes 
{\bf Z}_n$ in \cite{rozleigh} in a somewhat more 
complicated way.}. A simple way to see this for general 
$\Gamma$ is to note that
(in supersymmetric cases which we are focusing on here) an
individual twist with $d_a=2$ preserves ${\cal N}=2$ supersymmetry, so that
the corresponding gauge theory is anomaly free. This then immediately
implies that we do not have any non-Abelian gauge anomalies in a theory
with multiple such twists either. Indeed, such anomalies would have to
arise at the one-loop level. In the string theory language the relevant
diagram is a $g=0$, $b=2$ diagram (the annulus) with three external lines 
corresponding to non-Abelian gauge fields
attached to a single boundary. Thus, if we attach two external lines to
one boundary, while the third one to another boundary, the corresponding
diagram will vanish - indeed, the Chan-Paton structure of such a diagram
is given by ($\lambda_r$, $r=1,2,3$, are the Chan-Paton matrices
corresponding to the external lines)
\begin{equation}
 {\rm Tr}\left(\lambda_1\lambda_2\gamma_a\right){\rm Tr}\left(\lambda_3
 \gamma_a\right)~,
\end{equation}
which vanishes as for non-Abelian gauge fields
${\rm Tr}(\lambda_r)=0$, so ${\rm Tr}(\lambda_r\gamma_a)=0$ as well for by
definition $\lambda_r$ are invariant under the orbifold group action as
$\lambda_r$ correspond to the gauge bosons of the gauge group left unbroken
by the orbifold (in particular, note that $\lambda_r$ commute with
$\gamma_a$). As to the aforementioned diagram with all three external
lines attached to one boundary, its Chan-Paton structure is given by
\begin{equation}\label{anomaly}
 {\rm Tr}\left(\lambda_1\lambda_2\lambda_3\gamma_a\right)
 {\rm Tr}\left(\gamma_a\right)~.
\end{equation}
For twists with $d_a=2$, each diagram of this type is ${\cal N}=2$ 
supersymmetric as it does not contain any information about
further supersymmetry breaking (that is, the characters corresponding to the
world-sheet degrees of freedom multiplying the Chan-Paton trace (\ref{anomaly})
are those of an ${\cal N}=2$ theory), so such diagrams do not introduce
any anomalies. We therefore conclude that a theory with multiple $d_a=2$
twists with ${\rm Tr}(\gamma_a)\not=0$ is also anomaly free.

{}Thus, as we see, in supersymmetric cases we can have non-trivial
twists with $d_a=0,2$. Consistency of the background requires that
\begin{equation}
 {\rm Tr}(\gamma_a)=0~~{\rm for}~~d_a=0~,
\end{equation}
while for $d_a=2$ we do not have such a restriction.

\section{Non-conformal Large $N$ Gauge Theories}

{}In this section we discuss certain large $N$ gauge theories 
arising in the above 
setup with some twists $g_a$ with $d_a=2$ such that ${\rm Tr}(\gamma_a)\not=0$.
As we have already mentioned, such theories are not conformal.
The simplest examples of such theories are those with ${\cal N}=2$ 
supersymmetry. Perturbatively such theories are not renormalized beyond the
one-loop order. We can also construct non-conformal ${\cal N}=1$ 
supersymmetric theories in this way. In general such theories are rather 
complicated. However, as we will see in the following, certain non-trivial 
${\cal N}=1$ theories of this type have the property that in the large $N$
limit the leading (that is, planar) diagrams do not renormalize the
gauge theory correlators beyond the
one-loop order. In fact, the corresponding correlation functions in such an
${\cal N}=1$ theory are (up to overall numerical coefficients) the same as in  
a parent ${\cal N}=2$ theory. 

{}The simplest example of an ${\cal N}=2$ theory is obtained if we take 
${\cal M}={\bf C}^3/\Gamma$, where the action of $\Gamma\approx{\bf Z}_2$
on the complex coordinates $z_\alpha$ ($\alpha=1,2,3$)
on ${\cal M}$ is as follows: $R:z_1\rightarrow z_1$, $R:z_{2,3}
\rightarrow -z_{2,3}$ (where $R$ is the generator of $\Gamma$). Next, let us
choose the representation of $\Gamma$ when acting on the Chan-Paton charges
as follows: $\gamma_R=I_N$.
The massless spectrum of this model is the ${\cal N}=2$ supersymmetric
$U(N)$ gauge theory with no matter. The non-Abelian part of this theory is
not conformal, and is asymptotically free. As to the overall center-of-mass
$U(1)$, it is free, and can therefore be ignored. 

{}More generally, if we have twists with $\gamma_a\not=I_N$, the gauge group 
generically is a product of $U(N_k)$ factors, and we also have matter, which
can be obtained using the quiver diagrams 
(see \cite{dougmoore,JM,LNV}\footnote{A T-dual description of such models
can be studied in the context of the brane-box models \cite{HSU}.}). 
There is
always an overall center-of-mass $U(1)$, which is free. Other $U(1)$ factors,
however, run as the matter is charged under them. In the large $N$ limit,
however, these $U(1)$'s decouple in the infra-red, and can therefore also be 
ignored\footnote{As was pointed out in \cite{IRU,Poppitz}, in some cases these
$U(1)$'s are actually anomalous, and are broken at the tree-level via
the generalized Green-Schwarz mechanism. In particular, in cases with
twists with $d_a=0$ we have mixed $U(1)_k SU(N_l)^2$ anomalies. However, in 
the cases with matter we are interested in here we do have running $U(1)$'s 
that are anomaly free. These $U(1)$'s decouple in the infra-red
in the large $N$ limit.}. 

\subsection{Non-renormalization Theorems in ${\cal N}=1$ Theories}

{}As we have already mentioned, within the above setup we can construct
non-trivial ${\cal N}=1$ supersymmetric theories which are not renormalized
beyond the one-loop order in the planar limit.
The idea here is similar to that in \cite{z1}, where it was
noticed that in theories with vanishing twisted tadpoles the planar diagrams
are the same (up to overall numerical factors) as in the parent ${\cal N}=4$
theory. This is because the diagrams that contain information about 
supersymmetry breaking always contain twisted Chan-Paton traces 
${\rm Tr}(\gamma_a)$, which vanish in such theories. In this subsection we 
will consider theories where the planar diagrams are the same as in a parent
${\cal N}=2$ theory in essentially the same way. 

{}To obtain such models, consider an orbifold group $\Gamma$, which is a 
subgroup of $SU(3)$ but is not a subgroup of $SU(2)$. 
Let ${\widetilde \Gamma}$
be a non-trivial subgroup of $\Gamma$ such that ${\widetilde \Gamma}\subset
SU(2)$. We will allow the Chan-Paton matrices $\gamma_a$ corresponding to the
twists $g_a\in {\widetilde \Gamma}$ (which have $d_a=2$) not 
to be traceless, so that the corresponding
${\cal N}=2$ model is not conformal. However, we will require that
the other Chan-Paton matrices
$\gamma_a$ for the twists $g_a\not\in{\widetilde \Gamma}$ be traceless.
The resulting ${\cal N}=1$ model is not conformal. However, in the planar
limit the perturbative gauge theory amplitudes are not renormalized 
beyond one loop. 

{}The proof of this statement is straightforward. Thus, consider a planar
diagram with $b$ boundaries (but no 
handles)\footnote{Such a diagram corresponds to a $(b-1)$-loop diagram in the
field theory language.} 
with all external lines attached to a single boundary, which without
loss of generality can be chosen to be the outer boundary. 
Such a diagram is depicted in Fig.1. 
Next, we need to specify the twists on the boundaries. 
A convenient choice (consistent with that made for the annulus
amplitude (\ref{partition})) is given 
by\footnote{Here some care is needed in the 
cases where the orbifold group $\Gamma$ is non-Abelian, and we have to choose
base points on the world-sheet to define the twists. Our discussion here, 
however, is unmodified also in this case.}
\begin{equation}\label{mono}
 \gamma_{a_1}=\prod_{s=2}^{b} \gamma_{a_s}~,
\end{equation}
where $\gamma_{a_1}$ corresponds to the outer boundary, while 
$\gamma_{a_s}$, $s=2,\dots b$, correspond to inner boundaries.
Let $\lambda_r$, $r=1\dots M$, be the Chan-Paton matrices corresponding to the
external lines. Then the above planar diagram has the following
Chan-Paton group-theoretic dependence:
\begin{equation}
 \sum {\rm Tr}\left(\gamma_{a_1} \lambda_1\dots\lambda_M\right)
\prod_{s=2}^{b}  {\rm Tr}(\gamma_{a_s})~,
\end{equation}
where the sum involves all possible distributions of the $\gamma_{a_s}$ twists
that satisfy the condition (\ref{mono}), as well as permutations of the
$\lambda_r$ factors (note that $\lambda_r$ here correspond to the states
that are kept after the orbifold projections, so that they commute with 
all $\gamma_a$). Note that the 
diagrams with all twists $g_a\in{\widetilde \Gamma}$ are (up to overall
numerical coefficients) the same as in the parent ${\cal N}=2$ theory, and 
therefore vanish beyond one loop. All other diagrams contain at least one
twist $g_{a_s}\not\in {\widetilde \Gamma}$ 
with $s>1$. This follows from the
condition (\ref{mono}). This then implies that all such diagrams vanish as
\begin{equation}
 {\rm Tr}(\gamma_a)=0~,~~~g_a\not\in{\widetilde \Gamma}~.
\end{equation}
In particular, this implies that the non-Abelian gauge couplings do not run
in the large $N$ limit beyond one loop\footnote{More precisely, the
higher loop contributions to the gauge coupling running,
which come from the diagrams with handles, are subleading in the large $N$
limit compared with the leading one-loop contribution. This is analogous to 
what happens in theories discussed in \cite{KT}. In fact, the techniques
used there to prove that the higher loop corrections are subleading are very
similar to the one we are using here.}. 
The anomalous scaling dimensions (two-point functions corresponding to 
the wave-function renormalization) for matter fields\footnote{Note that in
such theories all matter fields
are non-trivially charged under the non-Abelian gauge subgroup.}, the 
corresponding diagrams are (up to overall numerical factors) the same as in the
parent ${\cal N}=2$ theory, so they are not renormalized in the planar
limit. Finally, the Yukawa (three-point) and quartic
scalar (four-point) couplings are not renormalized in perturbation theory
due to the ${\cal N}=1$ non-renormalization theorem for the superpotential.

{}Let us now consider a simple example of the aforementioned gauge theories.
Thus, let ${\widetilde 
\Gamma}=\{{\widetilde g}_a |a=1,\dots,|{\widetilde\Gamma}|\}\subset SU(2)$,
and let the corresponding twisted Chan-Paton matrices ${\widetilde \gamma}_a=
I_{2N}$. Let us assume that ${\widetilde g}_a$ act non-trivially on the 
complex coordinates $z_2,z_3$ on ${\widetilde {\cal M}}={\bf C}^3/{\widetilde
\Gamma}$, while leaving the coordinate $z_1$ untouched. Moreover, let us assume
that the orbifold group ${\widetilde \Gamma}$ is Abelian, and its action on the
coordinates $z_2,z_3$ is diagonal. Next, let $R$ be a generator of a 
${\bf Z}_2$ group with the following action: $R:z_{1,2}\rightarrow -z_{1,2}$,
$R:z_3\rightarrow z_3$. Note that $R$ commutes with ${\widetilde g}_a\in
{\widetilde \Gamma}$. The full orbifold group is given by $\Gamma=\{{\widetilde
g_a},{\widehat g}_a|a=1,\dots,|{\widetilde \Gamma}|\}$, which is also Abelian,
and ${\widehat g}_a\equiv R{\widetilde g}_a$. Let the twisted Chan-Paton 
matrix $\gamma_R={\rm diag}(I_N,-I_N)$. Then the theory is the ${\cal N}=1$
supersymmetric $U(N)\otimes U(N)$ gauge theory with chiral matter multiplets,
call them $\phi$ and $\chi$, in $({\bf N},{\overline {\bf N}})$ and
$({\overline{\bf N}},{\bf N})$, respectively (in the following we will ignore 
the $U(1)$ factors). Note that there is no tree-level superpotential in this
theory:
\begin{equation}
 {\cal W}_{\rm{\small tree}}=0~.
\end{equation} 
This implies that we can break the gauge group down to the diagonal $SU(N)$
subgroup by giving the appropriate vacuum expectation values to the fields
$\phi$ and $\chi$. The resulting theory in the IR is then the 
${\cal N}=2$ supersymmetric
$SU(N)$ gauge theory without matter. 
In the string theory language this corresponds to moving the 
D3-branes off the ${\bf Z}_2$ orbifold singularity, hence the ${\cal N}=2$
supersymmetry of the resulting theory.

\subsection{Pure Supergluodynamics}

{}Before we end this section, as an aside 
we would like to discuss an embedding of 
pure ${\cal N}=1$ supergluodynamics in the setup discussed in
section II. Thus, consider the orbifold group
$\Gamma\approx{\bf Z}_2\otimes {\bf Z}_2$ with the generators $R_1$ and
$R_2$ of the two ${\bf Z}_2$'s acting on the complex coordinates $z_\alpha$
on ${\cal M}={\bf C}^3/\Gamma$ as follows:
\begin{eqnarray}  
 R_\alpha : z_\beta\rightarrow -(-1)^{\delta_{\alpha\beta}} z_\beta~,
\end{eqnarray}
where $R_3\equiv R_1 R_2$. Next, let the twisted Chan-Paton
matrices be given by $\gamma_{R_\alpha}=I_N$. The D3-brane gauge theory is
then given by the ${\cal N}=1$ supersymmetric $SU(N)$ super-Yang-Mills theory
without matter (plus an overall $U(1)$ which can be ignored). 
Note that the real dimensions of the points fixed under 
the twists $R_\alpha$ are $d_\alpha=2$. 

{}The above embedding shows that at least perturbatively pure 
supergluodynamics possesses a discrete symmetry, which is not evident in the
field theory language. Thus, we have
the ${\bf Z}_2\otimes {\bf Z}_2$ symmetry as all the interactions involving
twisted closed string states must be invariant under the orbifold group.
In addition, we have a ${\bf Z}_3$ symmetry which permutes the closed string 
states coming from the $R_\alpha$ twisted sectors. In particular, the generator
$\theta$ of this ${\bf Z}_3$ group has the following action:
\begin{equation}
 \theta R_\alpha \theta^{-1}=R_{\alpha+1}~,~~~R_{\alpha+3}\equiv R_\alpha~.
\end{equation}
Thus, the ${\bf Z}_3$ subgroup does not commute with the ${\bf Z}_2\otimes
{\bf Z}_2$ subgroup. Thus, $R_1$ and $\theta$ generate a
non-Abelian discrete subgroup of $SO(3)$, namely, the 
{\em tetrahedral} group ${\cal T}$. 

{}Thus, as we see, the perturbative large $N$ pure superglue theory possesses
a discrete ${\bf Z}_3$ symmetry. In fact, this symmetry persists even at finite
$N$. Indeed, even the diagrams with handles possess this symmetry as this
symmetry is a symmetry of the underlying embedding of the pure 
supergluodynamics into string theory via the orbifold setup. In fact,
within this setup the ${\bf Z}_3$ symmetry is a discrete {\em gauge} symmetry.
Indeed, instead of ${\bf Z}_2\otimes {\bf Z}_2$, the orbifold group can be
chosen as the full tetrahedral group ${\cal T}$, where the generator $\theta$ 
of the ${\bf Z}_3$ subgroup of ${\cal T}$ acts on the complex coordinates 
$z_\alpha$ as follows: $\theta: z_\alpha\rightarrow z_{\alpha+1}$ 
($z_{\alpha+3}\equiv z_\alpha$). If we now choose 
the twisted Chan-Paton matrix $\gamma_\theta=I_N$, then the D3-brane gauge 
theory is still pure superglue. This embedding makes it evident that the
${\bf Z}_3$ symmetry is indeed a gauge symmetry, so it might be an exact 
symmetry of pure superglue even non-perturbatively\footnote{We can identify
other discrete gauge symmetries of this theory in a similar fashion. 
Thus, consider an orbifold group $\Gamma$ which is a subgroup
of $SO(3)$ but not a subgroup of $SU(2)$. The corresponding gauge theory has
${\cal N}=1$ supersymmetry. Note that the non-trivial twists in $\Gamma$ all
have $d_a=2$, and we can choose the corresponding twisted Chan-Paton matrices
as $\gamma_a=I_N$, so that (upon dropping the overall $U(1)$) 
we obtain the $SU(N)$ pure super-Yang-Mills theory. An appropriate Abelian
subgroup of $\Gamma$ is then a symmetry of this theory.}.

{}We would like to end this subsection by pointing out one immediate 
consequence of the aforementioned ${\bf Z}_3$ symmetry - both the one-loop as 
well as the two-loop\footnote{Note that beyond two loops the $\beta$-function 
becomes gauge dependent.} 
$\beta$-function coefficients in the pure supergluodynamics 
are multiples of 3.

\section{Large $N$ Limit and Brane-Bulk Duality}

{}As we mentioned in Introduction, perturbative expansion of
D3-brane gauge theories simplifies substantially in the large $N$ limit.
It is advantageous to work in the full string theory framework, 
which in a way is much simpler than the Feynman diagram techniques.
(At the end of the day we will take the $\alpha^\prime\rightarrow 0$ 
limit, which amounts to reducing the theory to the gauge theory subsector.)
Thus, in the string theory language there are two classes of diagrams 
we need to consider: ({\em i}) diagrams without handles;
({\em ii}) diagrams with handles. The latter correspond to closed string loops,
and can be neglected as they are subleading in the large $N$ limit. We will
therefore focus on the diagrams without handles. The latter diagrams can be
viewed as {\em tree-level} closed string graphs connecting various boundaries.
This suggests that (upon taking the $\alpha^\prime\rightarrow 0$ limit)
in some cases (which we will identify in a moment) we might be able to
rewrite the perturbative expansion 
in the large $N$ limit of the corresponding gauge theories in the language 
where various quantum corrections in the gauge theory are encoded in a {\em
classical} higher dimensional {\em field} theory. 
The effective quantum action for 
the four-dimensional large $N$ gauge theory is then obtained by 
starting with the corresponding higher dimensional classical action, 
integrating out the bulk fields,
and restricting to the gauge theory subsector\footnote{This way we can obtain
perturbative corrections to various {\em on-shell} gauge theory correlators
as we are using an on-shell formulation of the corresponding string theory.}. 
That is, we consider
the D-branes in the background of the bulk fields that is created by the 
D-branes themselves\footnote{Such a self-interaction in the case of strings
was discussed in \cite{DH,Ka}.}. The field theory cut-off in this language 
arises precisely from the fact that, in the above setup, 
the twisted closed string states have tadpoles resulting in logarithmic 
profiles in the extra dimensions, and we need
to regularize these profiles at the sources, that is, the D-branes, whose
locations in the extra dimensions are given by points in ${\cal M}$.

\subsection{Brane-Bulk Duality}

{}The aforementioned proposal that in the large $N$ limit 
various quantum corrections in the D-brane
gauge theory should in some cases be encoded in a higher dimensional classical
field theory is what we refer to as the {\em brane-bulk duality}. In this
subsection we would like to identify the cases where the brane-bulk duality
is applicable as stated above. Note that the brane-bulk duality is a 
consequence of the {\em open-closed duality} property of string theory. In 
particular, the diagrams without handles in the transverse closed string 
channel can indeed be viewed as tree-level closed string graphs connecting
various boundaries. At the one-loop order the corresponding diagram 
(up to external lines) is an
annulus, where two boundaries are connected by a single closed string tube.
In particular, this diagram does {\em not} involve closed string interactions.
Moreover, in the field theory limit $\alpha^\prime\rightarrow 0$ 
contributions due to the massive closed string states in the transverse closed
string channel are always finite. That is, these contributions correspond to
heavy string {\em thresholds}. In the field theory language they translate
into subtraction scheme dependent artifacts, which arise due to a particular
choice of regularization. Thus, at the one-loop order the contributions due to
massive string modes can be absorbed into the subtraction scheme dependence, so
that in the field theory limit the brane-bulk duality indeed holds - various
one-loop corrections in the gauge theory can be computed by calculating 
the classical D-brane self-interaction via the massless bulk fields.
 
{}Beyond the one-loop order, however, that is, when the number of boundaries
is greater than two, we must include closed string interactions in the 
transverse closed string channel. The massive closed string states are now
expected to contribute in a non-trivial way. In particular, {\em a priori}
there is no longer a clear interpretation of these contributions in terms of
thresholds. Another way of phrasing this is that at higher loop orders we can
have mixed contributions coming from both massless as well as massive states.
That is, in general the brane-bulk duality can still be formulated except that
at higher loop orders it will involve massless as well as {\em infinitely} 
many massive states as far as the corresponding higher dimensional classical
``field theory'' is concerned. In particular, in general this classical ``field
theory'' is nothing but the corresponding closed string theory, so we are back
to the original open-closed string duality.
 
{}There are, however, non-trivial cases where in the large $N$ limit the 
bulk closed string theory can be consistently truncated to a field theory 
containing only a finite number of massless fields. These cases are those where
in the planar limit the gauge theory perturbatively is not renormalized 
beyond one loop. Thus, this is clearly the case for ${\cal N}=2$ theories. In
theories without matter this, in fact, holds even at finite $N$. In theories
with matter, however, we have running $U(1)$'s whose decoupling is ensured
only in the large $N$ limit.

{}In subsection IIIA we discussed ${\cal N}=1$ theories which perturbatively
are not renormalized beyond the one-loop order in the planar limit. It is clear
that the aforementioned formulation of the brane-bulk duality also holds in
such theories, where, once again, various quantum corrections in the gauge 
theory are encoded in a higher dimensional classical field theory with
a finite number of massless fields. In fact, in ${\cal N}=2$ as well as 
${\cal N}=1$ theories the bulk fields entering the relevant part of the 
corresponding higher dimensional classical action are {\em twisted} closed
string states. Indeed, at the one-loop order the diagrams involving 
untwisted closed string states in the transverse closed string channel are
${\cal N}=4$ supersymmetric, so that they do not contribute into the 
renormalization of the corresponding (on-shell) gauge theory correlators.

{}To clarify the above discussion, let us give a schematic description of the
above procedure for obtaining quantum corrections in the corresponding 
gauge theories via the brane-bulk duality. 
Thus, we start from the classical action
\begin{equation}\label{action}
 S=S_{\rm{\small brane}}[\Phi]+S_{\rm{\small bulk}}[\sigma]+
 S_{\rm{\small int}}[\Phi,\sigma]~, 
\end{equation}
where $\Phi$ is a collective notation for the (fractional) 
D3-brane world-volume fields,
while $\sigma$ is a collective notation for the {\em massless twisted bosonic} 
(NS-NS and R-R) 
bulk fields; $S_{\rm{\small brane}}[\Phi]$ is the
classical four-dimensional action for the brane fields $\Phi$, while 
$S_{\rm{\small bulk}}[\sigma]$ is the classical higher dimensional action for
the bulk fields $\sigma$ (note that some of these fields such as twisted
closed string states propagate in less than 10 dimensions); finally, 
$S_{\rm{\small int}}[\Phi,\sigma]$, which is a classical
four-dimensional functional (with the integral over the (fractional) D3-brane 
world-volume), describes the coupling of the bulk fields 
to the brane fields, as well as to the brane itself
(in particular, it includes all tadpoles). Since here we are interested in
one-loop corrections in the gauge theory language, the action 
$S_{\rm{\small bulk}}[\sigma]$ contains only the terms quadratic in 
$\sigma$ (that is, the kinetic terms), while 
$S_{\rm{\small int}}[\Phi,\sigma]$  contains only the terms linear in $\sigma$.
Moreover, the fields 
$\sigma$ include only the twisted fields with non-vanishing
(logarithmic) tadpoles. 

{}Next, we solve the equations of motion for $\sigma$ following from the 
above action:
\begin{equation}
 {{\delta S_{\rm{\small bulk}}}\over {\delta\sigma}}=
 -{{\delta S_{\rm{\small int}}}\over {\delta\sigma}}~.
\end{equation} 
In particular, we are interested in the solution where the fields
$\sigma$ do {\em not} explicitly depend on the coordinates $x^\mu$ along the
D3-brane world-volume, but only functionally
via $\Phi(x^\mu),\partial_\nu\Phi(x^\mu),\dots$; this solution, however,
does in general explicitly depend on the coordinates $x^i$ transverse to 
the D-branes. Let us denote this 
solution via ${\overline\sigma}$. The effective quantum action for the 
fields $\Phi$ in the large $N$ limit is then given by
\begin{equation}\label{quantum}
 S^{\rm{\small qu}}_{\rm {\small brane }}[\Phi]=
 S_{\rm{\small brane}}[\Phi]+
 S_{\rm{\small int}}[\Phi,{\overline\sigma}]~.
\end{equation}
Note that $\left. {\overline\sigma}\right|_{\rm D3}$ are generically 
divergent, so we need to further clarify the meaning of the second term
in (\ref{quantum}). Thus, let 
\begin{equation}
 J\equiv \left.{{\delta S_{\rm{\small int}}}\over {\delta\sigma}}
 \right|_{\Phi=0}~.
\end{equation}
Note that $J$ are independent of $x^\mu$, and, in
fact, are nothing but the tadpoles for the fields $\sigma$. If all
${\rm Tr}(\gamma_a)=0$ (for $d_a=0,2$), then there are no tadpoles for the
corresponding twisted fields (that is, the corresponding
$J=0$), and the large $N$ gauge theory is finite 
\cite{z1}\footnote{More precisely, various $U(1)$
factors can still run, but, as we have already mentioned, 
in the large $N$ limit they decouple in the infra-red, and can therefore
be ignored.}. 
On the other hand,
if some ${\rm Tr}(\gamma_a)\not=0$ for twists with $d_a=2$, some divergences
no longer cancel, and we need to regularize 
$\left. {\overline\sigma}\right|_{\rm D3}$ in (\ref{quantum}). Note, however,
that the corresponding divergences are only logarithmic, and, in fact,
correspond to the running in the four-dimensional gauge theory, which is no
longer conformal.

\subsection{A Toy Model}

{}In this subsection we would like to address one technical point in the
context of the brane-bulk duality, namely, the issue of regularizing the
aforementioned divergences. Here we will consider
a simple toy model which possesses the main ingredients for
illustrating the regularization procedure. 
In fact, as we will see in the following sections, this model actually captures
all the key features of the ${\cal N}=2$ models as well the ${\cal N}=1$ 
models discussed in subsection IIIA.

{}Thus, consider the six dimensional theory with the following action:
\begin{equation}\label{toy}
 S=-{1\over 2}
 \int d^6 x~\partial_M\phi~ \partial^M \phi-a\int_\Sigma d^4 x~{\cal O}_4
 -b L^2 \int_\Sigma d^4x~\phi~{\cal O}_4-c L^{-2}\int_\Sigma d^4 x~\phi~.
\end{equation}
Here $\phi(x^M)$ is a six dimensional scalar field, whose dimension is
$({\rm mass})^2$; ${\cal O}_4(x^\mu)$ is a four-dimensional operator with
dimension $({\rm mass})^4$ localized on the hypersurface $\Sigma$;
the couplings $a,b,c$ (which are assumed to be non-vanishing)
are dimensionless; finally, $L$ is a parameter of dimension (length). 

{}In the above action $\phi$ is an analog of a twisted closed string state
with a non-vanishing tadpole (the last term in (\ref{toy})); $\Sigma$
plays the role of a 3-brane; ${\cal O}_4$ is an analog of a dimension-four
gauge theory operator such as ${\rm Tr}(F^2)$; finally, $L^2$ is analogous to
$\alpha^\prime$.

{}Following the procedure described in the previous subsection, we look for a
solution to the equation of motion for $\phi$: 
\begin{equation}
 \partial_M\partial^M\phi=\left[bL^2{\cal O}_4(y)+cL^{-2}\right]
 \delta^{(2)}(z)~.
\end{equation}
Here $z^i\equiv x^i$ are the coordinates transverse to the 3-brane (whose
location is chosen to be $z^i=0$); also, we use
the notation $y^\mu\equiv x^\mu$. It is convenient to Fourier transform to
the momentum space
\begin{equation}
 \phi(p,k)\equiv \int d^4 y ~d^2 z~e^{ip\cdot y+ik\cdot z}~
 \phi(y,z)~,
\end{equation} 
where $p^\mu$ and $k^i$ are the momenta corresponding to $y^\mu$ and $z^i$,
respectively. Thus, we have the following solution 
\begin{equation}
 {\overline \phi}(p,k)=-\left[bL^2{{\cal O}_4(p)\over
 {k^2+p^2}}+(2\pi)^4 cL^{-2}
 {\delta^{(4)}(p)\over k^2}\right]~. 
\end{equation}
This gives
\begin{equation}
 {\overline \phi}(p,z)=-\int {d^2 k\over (2\pi)^2}e^{-ik\cdot z}
 \left[bL^2{{\cal O}_4(p)\over
 {k^2+p^2}}+(2\pi)^4 cL^{-2}
 {\delta^{(4)}(p)\over k^2}\right]~.
\end{equation} 
The corresponding effective quantum action on the 3-brane is then given by
\begin{eqnarray}
 S^{\rm{\small qu}}_\Sigma&=&-a\int_\Sigma d^4 y~{\cal O}_4(y)
 -b L^2 \int_\Sigma d^4y~{\overline \phi}(y,z=0)~{\cal O}_4(y)-
 c L^{-2}\int_\Sigma d^4 y~{\overline\phi}(y,z=0)\nonumber\\
 &=&-a{\cal O}_4(p=0)-bL^2\int {d^4p\over (2\pi)^4}~
 {\overline\phi}(p,z=0)~{\cal O}_4(-p)-cL^{-2}{\overline\phi}(p=0,z=0)
 \nonumber\\
 \label{toyqu}
 &=&-\left[a-{bc\over 2\pi^2}\int {d^2 k\over k^2}\right]{\cal O}_4(p=0)
 +b^2 L^4\int {d^4 p\over(2\pi)^4}\int {d^2 k\over (2\pi)^2}
 {{\cal O}_4(p){\cal O}_4(-p)\over {k^2+p^2}}+\dots~,
\end{eqnarray} 
where the ellipses in the last line stand for 
a divergent constant piece (which does
not contain the operator ${\cal O}_4$, and is proportional to $c^2$ as it is
solely due to the presence of the $\phi$ tadpole). 

{}The second term in the last line in (\ref{toyqu}) is a non-local higher
dimensional operator, and it disappears in the $L^2\rightarrow 0$ limit.
The first term containing ${\cal O}_4(p=0)$ is the same as that in the 
classical 3-brane world-volume action except for the corresponding coupling -
the quantum coupling in $S^{\rm{\small qu}}_\Sigma$ is a renormalized
coupling:
\begin{equation}
 {\widetilde a}=a-{bc\over 2\pi^2}\int_{k^2=\mu^2}^{k^2=\Lambda^2}
 {d^2 k\over k^2}=
 a-{bc\over 2\pi}\ln\left({\Lambda^2\over\mu^2}\right)~. 
\end{equation}
Here $\Lambda$ is an ultra-violet (UV) cut-off, while $\mu$ is the infra-red
(IR) cut-off. In the four-dimensional 3-brane world-volume field theory
language the IR cut-off is interpreted as the renormalization group (RG)
scale at which the renormalized coupling ${\widetilde a}={\widetilde a}(\mu)$
is measured.

{}Thus, if in the above toy model we adapt the interpretation of the previous
subsection, from the classical dynamics of the bulk field $\phi$ we will obtain
the ``one-loop'' 
renormalization of the 3-brane world-volume field theory. In fact,
in this model the 3-brane world-volume theory is not renormalized beyond the
``one-loop'' order.

\section{Examples with ${\cal N}=2$ Supersymmetry}

{}In this section we would like to apply the brane-bulk duality discussed in 
the previous section to gauge theories arising in the orbifold construction
discussed in section II. Here we will focus on the simplest examples of this
type, in particular, those with ${\cal N}=2$ supersymmetry. Perturbatively such
gauge theories are not renormalized beyond the one-loop order even for 
finite $N$. Thus, the brane-bulk duality in such examples can be understood in
detail\footnote{Here we are considering the regime where the effective
't Hooft coupling is small. Supergravity duals of ${\cal N}=2$ theories in
the regime where the effective 't Hooft coupling is large were discussed
in \cite{JPP,Berto}.}.
 
{}We described the simplest example of an ${\cal N}=2$ gauge theory in the 
above context in the beginning of the previous section. In this example 
$\Gamma\approx {\bf Z}_2$, and the twisted Chan-Paton matrix $\gamma_R=I_N$.
The gauge group is $U(N)$, and we have no matter fields. In subsection A
we discuss the brane-bulk duality in this model in the case of unbroken
gauge symmetry. We will discuss the case of spontaneously broken gauge 
symmetry in subsection B. Finally, in subsection C we discuss examples with
matter.

\subsection{Unbroken Gauge Symmetry}

{}To discuss the brane-bulk duality in this model, we need the relevant 
part of the classical action containing the D3-brane as well as the bulk
fields. This action has the following form \cite{dougmoore} (we are not
including terms containing $F\wedge F$ for we are going to be interested 
in renormalization of the $F^2$ term)\footnote{Here we are using the
fact that $\gamma_R=I_N$. Also, in ${\rm Tr}\left[F_{\mu\nu}\right]$ only
the $U(1)$ contribution is non-vanishing.}:
\begin{eqnarray}
 S=&&-
 \int_{\rm D3} \Big(a~{\rm Tr}\left[F^{\mu\nu}F_{\mu\nu}\right]+
 b~\sigma~{\rm Tr}\left[F^{\mu\nu}F_{\mu\nu}\right]+
 c~\sigma+
 d~\epsilon^{\mu\nu\sigma\rho}~C_{\mu\nu}~
 {\rm Tr}\left[F_{\sigma\rho}\right]\Big)\nonumber\\
 &&-\int_{{\rm D3}\times {\bf R}^2} 
 \left({1\over 2}\partial^\mu\sigma~\partial_\mu\sigma+{1\over 12}
 H^{\mu\nu\sigma} H_{\mu\nu\sigma}\right)~.
\label{eq:actionu}
\end{eqnarray}
Here $\sigma$ is a twisted NS-NS scalar, while $C_{\mu\nu}$ is a twisted
two-form (whose field strength is $H_{\mu\nu\sigma}$). We have normalized
the kinetic terms of the bulk $\sigma$ and $C_{\mu\nu}$ fields in the standard
way. Once this normalization is fixed, the couplings $b,c,d$ can be 
determined. We discuss these couplings in Appendix A. The combinations
relevant for our discussion here are given by
\begin{equation}\label{bcd}
 bc=2Nd^2={N\over 8\pi}~.
\end{equation}  
Finally, the coupling $a$ is given by
\begin{equation}
 a={1\over 2 g_{\rm{\small YM}}^2}~, 
\end{equation}  
where $g_{\rm{\small YM}}$ is the tree-level Yang-Mills gauge coupling, and 
the $U(N)$ generators $T^A$ are normalized as $2{\rm Tr}\left(T^A T^B\right)=
\delta^{AB}$. The Yang-Mills gauge coupling is related to the string coupling
$g_s$ via $g^2_{\rm{\small YM}}=2\pi g_s$. 

{}As in section IV, we solve classical equations of motion for the
fields $\sigma$ and $C_{\mu\nu}$ (the latter is a gauge field, so
we must use gauge fixing such as $\partial^\mu C_{\mu\nu}=0$), 
and integrate them out of the 
classical action (\ref{eq:actionu}). The resulting effective quantum action
is given by (we are using the momentum representation, and drop
higher dimensional terms as well as those independent of $F_{\mu\nu}$):  
\begin{eqnarray}
 S^{\rm{\small qu}}_{\rm{\small brane}}=-\int {d^4 p\over (2\pi)^4}
 &&\left(\left[a-{bc\over 2\pi^2}\int {d^2 k\over k^2}\right]~
 {\rm Tr}\left[F^{\mu\nu}(p)~F_{\mu\nu}(-p)\right]+\right.\nonumber\\
 &&\left.{d^2\over \pi^2}\int {d^2 k\over{k^2+p^2}}~
 {\rm Tr}\left[F^{\mu\nu}(p)\right]~{\rm Tr}\left[F_{\mu\nu}(-p)
 \right]\right)~.
\end{eqnarray}
Since only the $U(1)$ subgroup contributes into ${\rm Tr}[F_{\mu\nu}]$,
we can rewrite the last term in the above expression in terms of the
$U(1)$ field strength $f_{\mu\nu}$ (note that the corresponding generator
is $I_N/\sqrt{2N}$): 
\begin{equation}
 {\rm Tr}[F^{\mu\nu}(p)]~{\rm Tr}[F_{\mu\nu}(-p)]=
 {\rm Tr}[f^{\mu\nu}(p)]~{\rm Tr}[f_{\mu\nu}(-p)]=
 N~{\rm Tr}[f^{\mu\nu}(p)~f_{\mu\nu}(-p)]~.
\end{equation}
Thus, we have
\begin{eqnarray}
 S^{\rm{\small qu}}_{\rm{\small brane}}=-\int {d^4 p\over (2\pi)^4}
 &&\left(\left[a-{bc\over 2\pi^2}\int {d^2 k\over k^2}\right]~
 {\rm Tr}\left[{\widehat F}^{\mu\nu}(p)~
 {\widehat F}_{\mu\nu}(-p)\right]+\right.\nonumber\\
 &&\left.\left[a-{bc\over 2\pi^2}\int {d^2 k\over k^2}+
 {N d^2\over \pi^2}\int {d^2 k\over{k^2+p^2}}\right]~
 {\rm Tr}\left[f^{\mu\nu}(p)~f_{\mu\nu}(-p)
 \right]\right)~,
\end{eqnarray}
where ${\widehat F}_{\mu\nu}$ is the $SU(N)$ field strength.
Note that the integrals contributing to the $f^2$ coupling individually are
UV divergent. However, since we have (\ref{bcd}), the total contribution is
UV finite. On the other hand, the first integral is IR divergent, while the
second integral is IR finite for $p^2>0$ (for this latter integral the $p^2$
term in the denominator plays the role of the IR cut-off). We must therefore
introduce an IR cut-off in the first integral. The fact that the 
$U(1)$ gauge coupling should not be renormalized then dictates that the IR
cut-off in the first integral must be chosen as follows\footnote{Here we 
should point out that the $U(1)$ gauge coupling can in general receive finite
(string) threshold corrections. Here, however, we ignore such corrections 
as we are interested in the field theory limit.}:
\begin{equation}
 a-{bc\over 2\pi^2}\int_{k^2=p^2} {d^2 k\over k^2}+
 {N d^2\over \pi^2}\int_{k^2=0} {d^2 k\over{k^2+p^2}}=a~.
\end{equation}
The effective quantum action therefore reads:
\begin{equation}
 S^{\rm{\small qu}}_{\rm{\small brane}}=-\int {d^4 p\over (2\pi)^4}
 \left({\widetilde a}(p^2)~
 {\rm Tr}\left[{\widehat F}^{\mu\nu}(p)~
 {\widehat F}_{\mu\nu}(-p)\right]+a~
 {\rm Tr}\left[f^{\mu\nu}(p)~f_{\mu\nu}(-p)
 \right]\right)~,
\end{equation}
where
\begin{equation}
 2{\widetilde a}(p^2)\equiv 2a-{bc\over \pi^2}\int_{k^2=p^2}^{k^2=\Lambda^2} 
 {d^2 k\over k^2}={1\over g_{\rm{\small YM}}^2}+
 {\beta_0\over 16\pi^2}~\ln\left(\Lambda^2\over p^2\right)~. 
\end{equation}
This is nothing but the one-loop renormalized Yang-Mills gauge coupling with
the $\beta$-function coefficient $\beta_0=-2N$.

\subsection{Spontaneously Broken Gauge Symmetry} 

{}In this subsection we will discuss the brane-bulk duality in the above
model in the case where the gauge symmetry is spontaneously broken:
$U(N)\rightarrow U(N_1)\otimes U(N_2)$, $N_1+N_2=N$. In the field theory
language this corresponds to the complex adjoint scalar in the ${\cal N}=2$
vector supermultiplet having an appropriate vacuum expectation value. In the
string theory language this corresponds to splitting the $N$ D3-branes into
two stacks of $N_1$ and $N_2$ D3-branes, and moving them apart by some distance
$X$ in the two real directions transverse to the D-branes untouched by the
${\bf Z}_2$ orbifold action. 

{}The relevant part of the classical action in this case is given by:
\begin{eqnarray}
 S=&&-\int_{{\rm D3}_1} a~{\rm Tr}[F_1^{\mu\nu}F_{1\mu\nu}]+
 b~\sigma~{\rm Tr}[F_1^{\mu\nu}F_{1\mu\nu}]+c_1~\sigma+
 d~\epsilon^{\mu\nu\sigma\rho}~C_{\mu\nu}~{\rm Tr}[F_{1\sigma\rho}]\nonumber\\
   &&-\int_{{\rm D3}_2} a~{\rm Tr}[F_2^{\mu\nu}F_{2\mu\nu}]+
 b~\sigma~{\rm Tr}[F_2^{\mu\nu}F_{2\mu\nu}]+c_2~\sigma+
 d~\epsilon^{\mu\nu\sigma\rho}~C_{\mu\nu}~{\rm Tr}[F_{2\sigma\rho}]\nonumber\\
   &&-\int_{{\rm D3}\times {\bf R}^2} 
 \left({1\over 2}\partial^\mu\sigma~\partial_\mu\sigma+{1\over 12}
 H^{\mu\nu\sigma} H_{\mu\nu\sigma}\right)~.
\label{eq:action}
\end{eqnarray}
The couplings $a,b,d$ are the same as before, while the relevant combinations
containing the couplings $c_{1,2}$ are now given by
\begin{equation}\label{couplings12}
 bc_{1,2}=2N_{1,2} d^2.
\end{equation}
In the following we will assume that the 
D3$_{1,2}$-branes (that is, the stacks of $N_{1,2}$ D3-branes) are located 
at $x^i=x^i_{1,2}$, respectively, in the extra two dimensions. Here $x^i$,
$i=1,2$, are real coordinates, and in the following we will use the notation 
$X^i\equiv x^i_1-x^i_2$, $X^2\equiv X^i X^i$.

{}Upon integrating out the bulk fields, we obtain:
\begin{eqnarray}
 S^{\rm{\small qu}}_{\rm{\small brane}}=&&-\int {d^4 p\over (2\pi)^4}
 \left(\left[a-{bc_1\over 2\pi^2}\int {d^2 k\over k^2}-
 {bc_2\over 2\pi^2}\int {d^2 k\over k^2}~e^{ik\cdot X}\right]~
 {\rm Tr}\left[F^{\mu\nu}_1(p)~F_{1\mu\nu}(-p)\right]+\right.\nonumber\\
 &&\left[a-{bc_2\over 2\pi^2}\int {d^2 k\over k^2}-
 {bc_1\over 2\pi^2}\int {d^2 k\over k^2}~e^{ik\cdot X}\right]~
 {\rm Tr}\left[F^{\mu\nu}_2(p)~F_{2\mu\nu}(-p)\right]+\nonumber\\
 &&{d^2\over \pi^2}\int {d^2 k\over{k^2+p^2}}~\Big(
 {\rm Tr}\left[F^{\mu\nu}_1(p)\right]~{\rm Tr}\left[F_{1\mu\nu}(-p)\right]+
 {\rm Tr}\left[F^{\mu\nu}_2(p)\right]~{\rm Tr}\left[F_{2\mu\nu}(-p)\right]
 \Big)+\nonumber\\
 &&\left. {2d^2\over \pi^2}\int {d^2 k\over{k^2+p^2}}~e^{ik\cdot X}~
 {\rm Tr}\left[F^{\mu\nu}_1(p)\right]~{\rm Tr}
 \left[F_{2\mu\nu}(-p)\right]\right)~.
\end{eqnarray}
As in the previous subsection, let us separate the $U(1)$ contributions:
\begin{eqnarray}
 &&S^{\rm{\small qu}}_{\rm{\small brane}}=-\int {d^4 p\over (2\pi)^4}
 \left(\left[a-{bc_1\over 2\pi^2}\int {d^2 k\over k^2}-
 {bc_2\over 2\pi^2}\int {d^2 k\over k^2}~e^{ik\cdot X}\right]~
 {\rm Tr}\left[{\widehat F}^{\mu\nu}_1(p)~{\widehat F}_{1\mu\nu}(-p)\right]+
 \right.\nonumber\\
 &&\left[a-{bc_2\over 2\pi^2}\int {d^2 k\over k^2}-
 {bc_1\over 2\pi^2}\int {d^2 k\over k^2}~e^{ik\cdot X}\right]~
 {\rm Tr}\left[{\widehat F}^{\mu\nu}_2(p)~{\widehat F}_{2\mu\nu}(-p)\right]
 +\nonumber\\
 &&\left[a-{bc_1\over 2\pi^2}\int {d^2 k\over k^2}-
 {bc_2\over 2\pi^2}\int {d^2 k\over k^2}~e^{ik\cdot X}+
 {N_1 d^2\over \pi^2}\int {d^2 k\over{k^2+p^2}}\right]~
 {\rm Tr}\left[f^{\mu\nu}_1(p)~f_{1\mu\nu}(-p)\right]
 +\nonumber\\
 &&\left[a-{bc_2\over 2\pi^2}\int {d^2 k\over k^2}-
 {bc_1\over 2\pi^2}\int {d^2 k\over k^2}~e^{ik\cdot X}+
 {N_2 d^2\over \pi^2}\int {d^2 k\over{k^2+p^2}}\right]~
 {\rm Tr}\left[f^{\mu\nu}_2(p)~f_{2\mu\nu}(-p)\right]
 +\nonumber\\
 &&\left. {2d^2\over \pi^2}\int {d^2 k\over{k^2+p^2}}~e^{ik\cdot X}~
 {\rm Tr}\left[f^{\mu\nu}_1(p)\right]~{\rm Tr}
 \left[f_{2\mu\nu}(-p)\right]\right)~.
\end{eqnarray}
To further simplify this expression, let us explicitly take the traces over
the $U(1)$ generators. Thus, we have
\begin{eqnarray}
 &&f_1^{\mu\nu}={1\over \sqrt{2N_1}}{\overline f}_1^{\mu\nu}{\rm diag}
 (I_{N_1},O_{N_2})~,\\
 &&f_2^{\mu\nu}={1\over \sqrt{2N_2}}{\overline f}_2^{\mu\nu}{\rm diag}
 (O_{N_1},I_{N_2})~,
\end{eqnarray} 
where $O_m$ is a null $m\times m$ matrix, and ${\overline f}_{1,2}^{\mu\nu}$
are the $U(1)$ field strengths. We therefore obtain:
\begin{eqnarray}
 &&S^{\rm{\small qu}}_{\rm{\small brane}}=-\int {d^4 p\over (2\pi)^4}
 \left(\left[a-{bc_1\over 2\pi^2}\int {d^2 k\over k^2}-
 {bc_2\over 2\pi^2}\int {d^2 k\over k^2}~e^{ik\cdot X}\right]~
 {\rm Tr}\left[{\widehat F}^{\mu\nu}_1(p)~{\widehat F}_{1\mu\nu}(-p)\right]+
 \right.\nonumber\\
 &&\left[a-{bc_2\over 2\pi^2}\int {d^2 k\over k^2}-
 {bc_1\over 2\pi^2}\int {d^2 k\over k^2}~e^{ik\cdot X}\right]~
 {\rm Tr}\left[{\widehat F}^{\mu\nu}_2(p)~{\widehat F}_{2\mu\nu}(-p)\right]
 +\nonumber\\
 &&\left[a-{bc_1\over 2\pi^2}\int {d^2 k\over k^2}-
 {bc_2\over 2\pi^2}\int {d^2 k\over k^2}~e^{ik\cdot X}+
 {N_1 d^2\over \pi^2}\int {d^2 k\over{k^2+p^2}}\right]~
 {1\over 2}{\overline f}^{\mu\nu}_1(p)~{\overline f}_{1\mu\nu}(-p)
 +\nonumber\\
 &&\left[a-{bc_2\over 2\pi^2}\int {d^2 k\over k^2}-
 {bc_1\over 2\pi^2}\int {d^2 k\over k^2}~e^{ik\cdot X}+
 {N_2 d^2\over \pi^2}\int {d^2 k\over{k^2+p^2}}\right]~
 {1\over 2}{\overline f}^{\mu\nu}_2(p)~{\overline f}_{2\mu\nu}(-p)
 +\nonumber\\
 &&\left. {\sqrt{N_1 N_2} d^2\over \pi^2}\int 
 {d^2 k\over{k^2+p^2}}~e^{ik\cdot X}~
 {\overline f}^{\mu\nu}_1(p)~
 {\overline f}_{2\mu\nu}(-p)\right)~.
\end{eqnarray}
Here it is convenient to rotate the above $U(1)$'s to the following basis
\begin{eqnarray}
 &&{\overline f}^{\mu\nu}_+\equiv \cos(\alpha) {\overline f}^{\mu\nu}_1 
 +\sin(\alpha) {\overline f}^{\mu\nu}_2~,\\
 &&{\overline f}^{\mu\nu}_-\equiv -\sin(\alpha) {\overline f}^{\mu\nu}_1 
 +\cos(\alpha) {\overline f}^{\mu\nu}_2~,
\end{eqnarray}
where
\begin{equation}
 \cos(\alpha)=\sqrt{{N_1\over N}}~,~~~\sin(\alpha)=\sqrt{{N_2\over N}}~.
\end{equation}
Here ${\overline f}^{\mu\nu}_{+}$ is the field strength of the overall
center-off-mass $U(1)$ in $U(N)=SU(N)\otimes U(1)_+$, while 
${\overline f}^{\mu\nu}_{-}$ is the field strength of the $U(1)$ in the
breaking $SU(N)\rightarrow SU(N_1)\otimes SU(N_2)\otimes U(1)_-$. Indeed,
we have
\begin{equation}
 f^{\mu\nu}_1+f^{\mu\nu}_2={1\over \sqrt{2N}}{\overline f}^{\mu\nu}_+
 {\rm diag}\left(I_{N_1},I_{N_2}\right)+{1\over \sqrt{2N}}
 {\overline f}^{\mu\nu}_-
 {\rm diag}\left(-\sqrt{N_2\over N_1}I_{N_1},\sqrt{N_1\over N_2} 
 I_{N_2}\right)~.
\end{equation}
The matrices multiplying ${\overline f}^{\mu\nu}_+$ 
and ${\overline f}^{\mu\nu}_-$ on the r.h.s.
of this equation are nothing but the properly normalized generators
of $U(1)_+$ and $U(1)_-$, respectively.

{}In the new basis we have  
\begin{eqnarray}
 &&\int{d^4 p\over (2\pi)^4}~
 \left(\left[a-{bc_1\over 2\pi^2}\int {d^2 k\over k^2}-
 {bc_2\over 2\pi^2}\int {d^2 k\over k^2}~e^{ik\cdot X}+
 {N_1 d^2\over \pi^2}\int {d^2 k\over{k^2+p^2}}\right]~
 {1\over 2}{\overline f}^{\mu\nu}_1(p)~{\overline f}_{1\mu\nu}(-p)
 +\right.\nonumber\\
 &&\left[a-{bc_2\over 2\pi^2}\int {d^2 k\over k^2}-
 {bc_1\over 2\pi^2}\int {d^2 k\over k^2}~e^{ik\cdot X}+
 {N_2 d^2\over \pi^2}\int {d^2 k\over{k^2+p^2}}\right]~
 {1\over 2}{\overline f}^{\mu\nu}_2(p)~{\overline f}_{2\mu\nu}(-p)
 +\nonumber\\
 &&\left.{\sqrt{N_1 N_2} d^2\over \pi^2}\int 
 {d^2 k\over{k^2+p^2}}~e^{ik\cdot X}~
 {\overline f}^{\mu\nu}_1(p)~
 {\overline f}_{2\mu\nu}(-p)\right)=\nonumber\\
 &&\int{d^4 p\over (2\pi)^4}~
 \left({\sqrt{N_1 N_2}d^2\over \pi^2}~{{N_1-N_2}\over N}~
 \left[\int{d^2 k\over {k^2+p^2}}~
 e^{ik\cdot X}-\int{d^2 k\over k^2}~e^{ik\cdot X}\right]~
 {\overline f}^{\mu\nu}_+(p)~{\overline f}_{-\mu\nu}(-p)+\right.\nonumber\\
 &&\left(a+{2N_1 N_2\over N}~{d^2\over \pi^2}~\left[\int{d^2 k\over {k^2+p^2}}~
 e^{ik\cdot X}-\int{d^2 k\over k^2}~e^{ik\cdot X}\right]\right)~
 {1\over 2}{\overline f}^{\mu\nu}_+(p)~{\overline f}_{+\mu\nu}(-p)+\nonumber\\
 &&\left.\left(a-{d^2\over N\pi^2}~\left[2N_1 N_2
 \int{d^2 k\over {k^2+p^2}}~
 e^{ik\cdot X}+\left[N_1^2+N_2^2\right]
 \int{d^2 k\over k^2}~e^{ik\cdot X}\right]\right)~
 {1\over 2}{\overline f}^{\mu\nu}_-(p)~{\overline f}_{-\mu\nu}(-p)\right)~.
 \label{U(1)s}
\end{eqnarray}
In arriving at this equation we have used (\ref{couplings12}), as well as the
fact that, as we discussed in the previous subsection, in the integral 
\begin{equation}\label{k^2}
 \int{d^2 k\over k^2}
\end{equation}
the IR cut-off is given by $k^2=p^2$, so that
\begin{equation}
 {bc_{1,2}\over 2\pi^2}\int_{k^2=p^2} {d^2 k\over k^2}-{N_{1,2}d^2\over\pi^2}
 \int {d^2 k\over{k^2+p^2}}=0~.
\end{equation}
In fact, the r.h.s. of (\ref{U(1)s}) further simplifies once we go to the
field theory limit. This limit is given by
\begin{equation}
 \alpha^\prime\rightarrow 0~,~~~X^i\rightarrow 0~,~~~\phi^i\equiv {X^i\over
 \alpha^\prime}={\rm fixed}~.
\end{equation}
In this limit the two stacks of D3-branes come on top of each other, but the
gauge symmetry is still $U(N_1)\otimes U(N_2)$ - the original $U(N)$ gauge
group is broken by the adjoint Higgs vacuum expectation value parametrized by
$\phi^i$. In particular, in this limit we have
\begin{equation}\label{vanish}
 \int{d^2 k\over {k^2+p^2}}~
 e^{ik\cdot X}-\int
 {d^2 k\over k^2}~e^{ik\cdot X}\rightarrow 0~.
\end{equation} 
This implies that on the r.h.s. of (\ref{U(1)s}) the 
${\overline f}^{\mu\nu}_+~
{\overline f}_{-\mu\nu}$ term goes to zero, the coupling for the 
${\overline f}^{\mu\nu}_+~ {\overline f}_{+\mu\nu}$ 
term goes to $a/2$ (that is, this coupling is not renormalized,
which is consistent with the fact that the overall center-of-mass $U(1)$
should not run), while the coupling for the ${\overline f}^{\mu\nu}_-~
{\overline f}_{-\mu\nu}$ term is renormalized as follows:
\begin{equation}\label{U(1)_-}
 {\widetilde a}_-(p^2,\phi^2)=
 a-{Nd^2\over\pi^2}\int {d^2 k\over k^2}~e^{ik\cdot X}~.
\end{equation}
This coupling, as well as the non-Abelian gauge couplings, need to be
regularized. This regularization, however, depends on whether $p^2\gg
\phi^2$ or $p^2\ll \phi^2$. 
This is because the aforementioned $\alpha^\prime\rightarrow 0$ limit
must be taken differently depending on $p^2$. However, (\ref{vanish}) holds
regardless of how the limit is taken.  

{}Collecting the above results we see that the effective quantum action is
given by:
\begin{eqnarray}
 S^{\rm{\small qu}}_{\rm{\small brane}}=-\int {d^4 p\over (2\pi)^4}
 \Big(&&{\widetilde a}_1(p^2,\phi^2)~
 {\rm Tr}\left[{\widehat F}^{\mu\nu}_1(p)~{\widehat F}_{1\mu\nu}(-p)\right]+
 {\widetilde a}_2(p^2,\phi^2)~
 {\rm Tr}\left[{\widehat F}^{\mu\nu}_2(p)~{\widehat F}_{2\mu\nu}(-p)\right]
 +\nonumber\\
 &&{1\over 2} a~{\overline f}^{\mu\nu}_+(p)~{\overline f}_{+\mu\nu}(-p)
 +{1\over 2}{\widetilde a}_-(p^2,\phi^2)~
 {\overline f}^{\mu\nu}_-(p)~{\overline f}_{-\mu\nu}(-p)\Big)~, 
\end{eqnarray}
where the $U(1)_-$ coupling is given by (\ref{U(1)_-}), while the non-Abelian
couplings are given by:
\begin{eqnarray}
 &&{\widetilde a}_1(p^2,\phi^2)=a-{bc_1\over 2\pi^2}\int {d^2 k\over k^2}-
 {bc_2\over 2\pi^2}\int {d^2 k\over k^2}~e^{ik\cdot X}~,\\
 &&{\widetilde a}_2(p^2,\phi^2)=a-{bc_2\over 2\pi^2}\int {d^2 k\over k^2}-
 {bc_1\over 2\pi^2}\int {d^2 k\over k^2}~e^{ik\cdot X}~.
\end{eqnarray}
As before, the integral (\ref{k^2}) is regularized as follows:
\begin{equation}\label{k^2reg}
 \int_{k^2=p^2}^{k^2=\Lambda^2} {d^2 k\over k^2}=\pi~\ln\left({\Lambda^2
 \over p^2}\right)~.
\end{equation}
However, as we have already mentioned, the regularization of the integral
\begin{equation}\label{k^2X}
 \int {d^2 k\over k^2}~e^{ik\cdot X}
\end{equation}
depends upon $p^2$. For $p^2\gg \phi^2$ we take the $\alpha^\prime\rightarrow
0$ limit in the exponent
\begin{equation}
 \int {d^2 k\over k^2}~e^{ik\cdot X}\rightarrow \int {d^2 k\over k^2}~,
\end{equation}
and regularize the resulting integral as in (\ref{k^2reg}). For $p^2\gg \phi^2$
we therefore have
\begin{equation}
 2{\widetilde a}_1=2{\widetilde a}_2=2{\widetilde a}_-={1\over 
 g^2_{\rm {\small YM}}}+{\beta_0\over 16\pi^2}~
 \ln\left({\Lambda^2\over p^2}\right)~.
\end{equation}
The r.h.s. of this equation is nothing but the one-loop renormalized $SU(N)$
gauge coupling with the $\beta$-function coefficient $\beta_0=-2N$. All three
gauge couplings, that is, those for $SU(N_1)$, $SU(N_2)$ and $U(1)_-$, run
together at $p^2\gg \phi^2$ as at large momenta the effects of the $SU(N)
\rightarrow SU(N_1)\otimes SU(N_2)\otimes U(1)_-$ breaking are negligible.

{}Now, at small momenta $p^2\ll \phi^2$ the $\alpha^\prime\rightarrow 0$ limit
must be taken differently. In particular, in the integral (\ref{k^2X}) we 
first redefine the integration variables via $\zeta^i\equiv \alpha^\prime k^i$.
The resulting integral 
\begin{equation}\label{zeta^2}
 \int {d^2\zeta\over \zeta^2}~e^{i\zeta\cdot\phi}
\end{equation}
must then be regularized for small $\zeta^2$. Note that $\zeta^2$ has dimension
of $({\rm length})^2$, so this is a UV divergence. The regularized integral
(\ref{zeta^2}) is then given by
\begin{equation}
 \pi~\ln\left({\xi^2\Lambda^2\over\phi^2}\right)~,
\end{equation}
where $\xi$ parametrizes the subtraction scheme dependence (see below). For
small momenta $p^2\ll\phi^2$ we therefore have:
\begin{eqnarray}
 &&2{\widetilde a}_1={1\over 
 g^2_{\rm {\small YM}}}+{\beta^{(1)}_0\over 16\pi^2}~
 \ln\left({\Lambda^2\over p^2}\right)+
 {\beta^{(2)}_0\over 16\pi^2}~\ln\left({\xi^2\Lambda^2\over \phi^2}\right)~,\\ 
 &&2{\widetilde a}_2={1\over 
 g^2_{\rm {\small YM}}}+{\beta^{(2)}_0\over 16\pi^2}~
 \ln\left({\Lambda^2\over p^2}\right)+
 {\beta^{(1)}_0\over 16\pi^2}~\ln\left({\xi^2\Lambda^2\over \phi^2}\right)~,\\
 &&2{\widetilde a}_-={1\over g^2_{\rm {\small YM}}}+
 {\beta_0\over 16\pi^2}~\ln\left({\xi^2\Lambda^2\over \phi^2}\right)~,
\end{eqnarray}
where $\beta^{(1)}_0=-2N_1$ and $\beta^{(2)}_0=-2N_2$ are the one-loop beta
function coefficients for $SU(N_1)$ and $SU(N_2)$, respectively. In the above
expressions the terms proportional to
$\ln\left({\xi^2\Lambda^2/\phi^2}\right)$ correspond to the {\em threshold}
corrections due to the massive gauge bosons (that is, the gauge bosons that
become heavy in the Higgs mechanism).

{}As usual, to connect the gauge coupling evolution above and below the 
threshold, we need to specify a subtraction scheme. Thus, we can choose the
subtraction scheme where the gauge couplings are matched at the scale $p^2=
M^2$, where $M$ is the mass of the heavy gauge bosons. In particular, this
implies that
\begin{equation}
 \xi^2={\phi^2\over M^2}~. 
\end{equation}
We then have the
following gauge coupling running. For $p^2\geq M^2$
\begin{equation}
 2{\widetilde a}_1=2{\widetilde a}_2=2{\widetilde a}_-= {1\over 
 {\widetilde g}^2_{\rm {\small YM}}(p^2)}~,
\end{equation}
where
\begin{equation}
 {1\over {\widetilde g}^2_{\rm {\small YM}}(p^2)}\equiv {1\over 
 g^2_{\rm {\small YM}}}+{\beta_0\over 16\pi^2}~
 \ln\left({\Lambda^2\over p^2}\right)~,
\end{equation}
which is the $SU(N)$ gauge coupling at $p^2\geq M^2$. For $p^2<M^2$ we have
\begin{eqnarray}
 &&2{\widetilde a}_1={1\over 
 {\widetilde g}^2_{\rm {\small YM}}(M^2)}+{\beta^{(1)}_0\over 16\pi^2}~
 \ln\left({M^2\over p^2}\right)~,\\ 
 &&2{\widetilde a}_2={1\over 
 {\widetilde g}^2_{\rm {\small YM}}(M^2)}+{\beta^{(2)}_0\over 16\pi^2}~
 \ln\left({M^2\over p^2}\right)~,\\
 &&2{\widetilde a}_-={1\over {\widetilde g}^2_{\rm {\small YM}}(M^2)}~,
\end{eqnarray}
so that below the threshold the $SU(N_1)$ and $SU(N_2)$ gauge couplings run
with the $\beta$-function coefficients 
$\beta^{(1)}_0=-2N_1$ and $\beta^{(2)}_0=-2N_2$, respectively, while the
$U(1)_-$ gauge coupling does not run at all.

{}Thus, using the brane-bulk duality approach we reproduce the expected 
perturbative running of gauge couplings in the corresponding ${\cal N}=2$ gauge
theories. In the brane-bulk duality approach, however, we do not perform any
loop computations. Rather, the information about the loop corrections in 
gauge theory is encoded in the corresponding classical higher dimensional field
theory. This is a consequence of the fact that the brane-bulk duality simply 
follows from the closed-open string duality. 

{}Before we end this subsection, the following remark is in order. In the above
computations we had to regularize various (logarithmically) divergent 
integrals. The corresponding regularizations depend upon the four-dimensional
momentum squared. For instance, in this subsection we saw that the 
regularization of the integrals containing the information about the 
threshold depends on whether $p^2$ is above or below the threshold. In 
particular, the details of the corresponding regularization are somewhat 
different from what happens in the direct loop computation in the field theory
language. This appears to be a common feature of string theory 
embeddings of gauge theories as far as computations of, say, gauge coupling
running are concerned. In particular, to reproduce the gauge coupling running, 
it appears to be necessary to introduce an 
(IR) cut-off which is $p^2$ dependent. This
appears to be a consequence of the fact that such computations are
typically done within the {\em on-shell} formulation of string 
theory\footnote{We would like to thank Tom Taylor for a discussion on this
issue.}.

\subsection{Cases with Matter}

{}For completeness in this subsection we would like to briefly discuss 
examples of ${\cal N}=2$ gauge theories with matter. The simplest examples of
this type are obtained via the aforementioned ${\bf Z}_2$ orbifold 
construction with the twisted Chan-Paton matrix given by $\gamma_R=
{\rm diag}(I_{N_1},-I_{N_2})$, where $N_1+N_2=N$. In this case the gauge group
is $U(N_1)\otimes U(N_2)$, and we have matter hypermultiplets in $({\bf N}_1,
{\overline {\bf N}}_2)(+1,-1)$ and $({\overline {\bf N}}_1,
{\bf N}_2)(-1,+1)$, where the $U(1)$ charges are given in parentheses. 
Note that all $N$ D3-branes are now coincident, and the gauge 
symmetry is broken due to the non-trivial orbifold action on the Chan-Paton 
factors.

{}The relevant part of the classical action is given by (for simplicity
the matter field contributions are not shown, and we use the notation
$F^{\mu\nu}={\rm diag}(F^{\mu\nu}_1,F^{\mu\nu}_2)$):
\begin{eqnarray}
 S=&&-
 \int_{\rm D3} \Big(a~{\rm Tr}\left[F^{\mu\nu}F_{\mu\nu}\right]+
 b~\sigma~{\rm Tr}\left[\gamma_R~F^{\mu\nu}F_{\mu\nu}\right]+
 {\rm Tr}(\gamma_R)~{\widehat c}~\sigma+
 d~\epsilon^{\mu\nu\sigma\rho}~C_{\mu\nu}~
 {\rm Tr}\left[\gamma_R~F_{\sigma\rho}\right]\Big)\nonumber\\
 &&-\int_{{\rm D3}\times {\bf R}^2} 
 \left({1\over 2}\partial^\mu\sigma~\partial_\mu\sigma+{1\over 12}
 H^{\mu\nu\sigma} H_{\mu\nu\sigma}\right)=\nonumber\\
 &&-
 \int_{\rm D3} \Big(a~\left[{\rm Tr}\Big[F^{\mu\nu}_1 F_{1\mu\nu}\right]+
 {\rm Tr}\left[F^{\mu\nu}_2 F_{2\mu\nu}\right]\Big]+
 b~\sigma~\Big[{\rm Tr}\left[F^{\mu\nu}_1 F_{1\mu\nu}\right]-
 {\rm Tr}\left[F^{\mu\nu}_2 F_{2\mu\nu}\right]\Big]\nonumber\\
 &&+(N_1-N_2)~{\widehat c}~\sigma+
 d~\epsilon^{\mu\nu\sigma\rho}~C_{\mu\nu}~
 \Big[{\rm Tr}\left[F_{1\sigma\rho}\right]-
 {\rm Tr}\left[F_{2\sigma\rho}\right]\Big]\Big)\nonumber\\
 &&-\int_{{\rm D3}\times {\bf R}^2} 
 \left({1\over 2}\partial^\mu\sigma~\partial_\mu\sigma+{1\over 12}
 H^{\mu\nu\sigma} H_{\mu\nu\sigma}\right)~,
\label{eq:actionu2}
\end{eqnarray}
where the couplings $a,b,d$ are the same as before, while
\begin{equation}
 b{\widehat c}=2d^2~.
\end{equation}
Note that for $N_1=0$ or $N_2=0$ we recover the action (\ref{eq:actionu}). 

{}Integrating out the bulk fields, we obtain:
\begin{eqnarray}
 &&S^{\rm{\small qu}}_{\rm{\small brane}}=-\int {d^4 p\over (2\pi)^4}
 \left(\left[a-{b{\widehat c}\over 2\pi^2}~(N_1-N_2)\int {d^2 k\over k^2}
 \right]~
 {\rm Tr}\left[F^{\mu\nu}_1(p)~F_{1\mu\nu}(-p)\right]+
 \right.\nonumber\\
 &&\left[a-{b{\widehat c}\over 2\pi^2}~(N_2-N_1)\int {d^2 k\over k^2}
 \right]~
 {\rm Tr}\left[F^{\mu\nu}_2(p)~F_{2\mu\nu}(-p)\right]+\nonumber\\
 &&\left.{d^2\over \pi^2}\int {d^2 k\over{k^2+p^2}}~
 \left({\rm Tr}[F^{\mu\nu}_1(p)]-{\rm Tr}[F^{\mu\nu}_2(p)]\right)~
 \left({\rm Tr}[F_{1\mu\nu}(-p)]-{\rm Tr}[F_{2\mu\nu}(-p)]\right)\right)~.
\end{eqnarray}
The $U(1)$ contributions can be extracted as in the previous subsection,
and, in fact, all the corresponding normalizations are exactly the same as 
before. We therefore obtain:
\begin{eqnarray}
 S^{\rm{\small qu}}_{\rm{\small brane}}=-\int {d^4 p\over (2\pi)^4}
 \Big(&&{\widetilde a}_1(p^2)~
 {\rm Tr}\left[{\widehat F}^{\mu\nu}_1(p)~{\widehat F}_{1\mu\nu}(-p)\right]+
 {\widetilde a}_2(p^2)~
 {\rm Tr}\left[{\widehat F}^{\mu\nu}_2(p)~{\widehat F}_{2\mu\nu}(-p)\right]
 +\nonumber\\
 &&{1\over 2} a~{\overline f}^{\mu\nu}_+(p)~{\overline f}_{+\mu\nu}(-p)
 +{1\over 2}{\widetilde a}_-(p^2)~
 {\overline f}^{\mu\nu}_-(p)~{\overline f}_{-\mu\nu}(-p)\Big)~, 
\end{eqnarray}
where the $U(1)_-$ coupling is given by 
\begin{equation}
 2{\widetilde a}_-={1\over g^2_{\rm {\small YM}}}+
 {\beta^{(-)}_0\over 16\pi^2}~\ln\left(
 {\Lambda^2\over p^2}\right)~,
\end{equation}
while the non-Abelian couplings are given by
\begin{equation}
 2{\widetilde a}_{1,2}={1\over g^2_{\rm {\small YM}}}+
 {\beta^{(1,2)}_0\over 16\pi^2}~\ln\left(
 {\Lambda^2\over p^2}\right)~,
\end{equation}
where $\beta^{(-)}_0=2N$ is the $U(1)_-$ one-loop $\beta$-function coefficient,
while $\beta^{(1)}_0=-2(N_1-N_2)$ and $\beta^{(2)}_0=-2(N_2-N_1)$ are the
one-loop $\beta$-function coefficients for the $SU(N_1)$ and $SU(N_2)$
subgroups, respectively. Thus, as we see, once again, 
we correctly reproduce the running of the
gauge couplings for the $SU(N_1)\otimes SU(N_2)\otimes U(1)_-$ subgroup
(the overall center-of-mass $U(1)_+$ does not run as there is no matter 
charged under it). 

{}Note that the twisted tadpole for $\sigma$ vanishes for $N_1=N_2$, that is,
for ${\rm Tr}(\gamma_R)=0$. The non-Abelian one-loop 
$\beta$-function coefficients in
this case vanish. The $U(1)_-$ one-loop $\beta$-function coefficient, however,
is still non-vanishing, so the $U(1)_-$ gauge coupling runs. In the large $N$
limit $U(1)_-$ decouples in the IR, and we are left with an ${\cal N}=2$
superconformal field theory. 

\section{Examples with ${\cal N}=1$ Supersymmetry}

{}In this section we would like to discuss the brane-bulk duality in ${\cal 
N}=1$ supersymmetric theories of the type discussed in subsection IIIA.
Thus, let us consider the example where $\Gamma={\bf Z}_2\otimes {\bf Z}_2$, 
and the action of the generators $R_1$ and $R_2$ of the two ${\bf Z}_2$'s on
the complex coordinates $z_\alpha$ on ${\cal M}={\cal C}^3/\Gamma$ is given 
by (here $R_3\equiv R_1 R_2$)
\begin{equation}
 R_\alpha: -z_\beta \rightarrow (-1)^{\delta_{\alpha\beta}} z_\beta~.
\end{equation} 
Let us choose the twisted Chan-Paton matrices as follows: $\gamma_{R_1}=
{\rm diag}(I_{N_1},I_{N_2})$, $\gamma_{R_2}=\gamma_{R_3}=
{\rm diag}(I_{N_1},-I_{N_2})$, where $N_1+N_2=N$. In this case we have the
${\cal N}=1$ supersymmetric $U(N_1)\otimes U(N_2)$ gauge theory with 
chiral matter in $({\bf N}_1,
{\overline {\bf N}}_2)(+1,-1)$ and $({\overline {\bf N}}_1,
{\bf N}_2)(-1,+1)$, where the $U(1)$ charges are given in parentheses. 
Note that all $N$ D3-branes are coincident, and the gauge 
symmetry is broken due to the non-trivial orbifold action on the Chan-Paton 
factors. For $N_1=N_2$ we have an ${\cal N}=1$ theory of the type discussed in
subsection IIIA. In particular, in the planar limit the correlation functions 
in this theory are the same as in the parent ${\cal N}=2$ supersymmetric gauge
theory with $U(N)$ gauge group and no matter. We can therefore discuss this
theory in the context of the brane-bulk duality as in the large $N$ limit we 
have the corresponding non-renormalization theorem beyond the one-loop order.

{}For calculational convenience in the following we will keep $N_1$ and $N_2$
arbitrary. The calculation of the one-loop effective quantum action then gives
a correct result even for $N_1\not= N_2$, but only for $N_1=N_2$ do we have
the non-renormalization theorem beyond the one-loop order.
The relevant part of the classical action is given by (for simplicity
the matter field contributions are not shown, and we use the notation
$F^{\mu\nu}={\rm diag}(F^{\mu\nu}_1,F^{\mu\nu}_2)$):
\begin{eqnarray}
 S=&&-
 \int_{\rm D3} \Big(a~{\rm Tr}\left[F^{\mu\nu}F_{\mu\nu}\right]\nonumber\\
 &&+{1\over\sqrt{2}}
 \sum_\alpha\Big[b~\sigma_\alpha~{\rm Tr}\left[\gamma_{R_\alpha}~F^{\mu\nu}
 F_{\mu\nu}\right]+
 {\rm Tr}(\gamma_{R_\alpha})~{\widehat c}~\sigma_\alpha+
 d~\epsilon^{\mu\nu\sigma\rho}~C_{\alpha\mu\nu}~
 {\rm Tr}\left[\gamma_{R_\alpha}~F_{\sigma\rho}\right]\Big]\Big)\nonumber\\
 &&-\sum_\alpha \int_{{\rm D3}\times {\cal F}_\alpha} 
 \left({1\over 2}\partial^\mu\sigma_\alpha~\partial_\mu\sigma_\alpha+
 {1\over 12}
 H_\alpha^{\mu\nu\sigma} H_{\alpha\mu\nu\sigma}\right)~,
\label{eq:actionu3}
\end{eqnarray}
where the couplings $a,b,{\widehat c},d$ are the same as before, and the 
overall factor of $1/\sqrt{2}$ is due to the fact that the $R_\alpha$ 
twisted fields $\sigma_\alpha$ and $C^{\mu\nu}_\alpha$ now propagate in
${\bf R}^{1,3}\times {\cal F}_\alpha$, where each fixed point set 
${\cal F}_\alpha$ is an orbifold ${\bf R}^2/{\bf Z}_2$.

{}Integrating out the bulk fields, we obtain:
\begin{eqnarray}
 &&S^{\rm{\small qu}}_{\rm{\small brane}}=-\int {d^4 p\over (2\pi)^4}
 \left(\left[a-{b{\widehat c}\over 4\pi^2}~(3N_1-N_2)\int {d^2 k\over k^2}
 \right]~
 {\rm Tr}\left[F^{\mu\nu}_1(p)~F_{1\mu\nu}(-p)\right]+
 \right.\nonumber\\
 &&\left[a-{b{\widehat c}\over 4\pi^2}~(3N_2-N_1)\int {d^2 k\over k^2}
 \right]~
 {\rm Tr}\left[F^{\mu\nu}_2(p)~F_{2\mu\nu}(-p)\right]+\nonumber\\ 
 &&{d^2\over 2\pi^2}\int {d^2 k\over{k^2+p^2}}~
 \left({\rm Tr}[F^{\mu\nu}_1(p)]+{\rm Tr}[F^{\mu\nu}_2(p)]\right)~
 \left({\rm Tr}[F_{1\mu\nu}(-p)]+{\rm Tr}[F_{2\mu\nu}(-p)]\right)\nonumber\\
 &&\left.{d^2\over \pi^2}\int {d^2 k\over{k^2+p^2}}~
 \left({\rm Tr}[F^{\mu\nu}_1(p)]-{\rm Tr}[F^{\mu\nu}_2(p)]\right)~
 \left({\rm Tr}[F_{1\mu\nu}(-p)]-{\rm Tr}[F_{2\mu\nu}(-p)]\right)\right)~.
\end{eqnarray}
The $U(1)$ contributions can be extracted as in the previous section,
and, in fact, all the corresponding normalizations are exactly the same as 
before. We therefore obtain:
\begin{eqnarray}
 S^{\rm{\small qu}}_{\rm{\small brane}}=-\int {d^4 p\over (2\pi)^4}
 \Big(&&{\widetilde a}_1(p^2)~
 {\rm Tr}\left[{\widehat F}^{\mu\nu}_1(p)~{\widehat F}_{1\mu\nu}(-p)\right]+
 {\widetilde a}_2(p^2)~
 {\rm Tr}\left[{\widehat F}^{\mu\nu}_2(p)~{\widehat F}_{2\mu\nu}(-p)\right]
 +\nonumber\\
 &&{1\over 2} a~{\overline f}^{\mu\nu}_+(p)~{\overline f}_{+\mu\nu}(-p)
 +{1\over 2}{\widetilde a}_-(p^2)~
 {\overline f}^{\mu\nu}_-(p)~{\overline f}_{-\mu\nu}(-p)\Big)~, 
\end{eqnarray}
where the $U(1)_-$ coupling is given by 
\begin{equation}
 2{\widetilde a}_-={1\over g^2_{\rm {\small YM}}}+
 {\beta^{(-)}_0\over 16\pi^2}~\ln\left(
 {\Lambda^2\over p^2}\right)~,
\end{equation}
while the non-Abelian couplings are given by
\begin{equation}
 2{\widetilde a}_{1,2}={1\over g^2_{\rm {\small YM}}}+
 {\beta^{(1,2)}_0\over 16\pi^2}~\ln\left(
 {\Lambda^2\over p^2}\right)~,
\end{equation}
where $\beta^{(-)}_0=N$ is the $U(1)_-$ one-loop $\beta$-function coefficient,
while $\beta^{(1)}_0=-(3N_1-N_2)$ and $\beta^{(2)}_0=-(3N_2-N_1)$ are the
one-loop $\beta$-function coefficients for the $SU(N_1)$ and $SU(N_2)$
subgroups, respectively. Thus, we correctly reproduce the running of the
gauge couplings for the $SU(N_1)\otimes SU(N_2)\otimes U(1)_-$ subgroup
(the overall center-of-mass $U(1)_+$ does not run as there is no matter 
charged under it). 

{}Note that the twisted tadpole for $\sigma_1$ does not vanish, so that the
non-Abelian part of the gauge theory is non-conformal even for $N_1=N_2\equiv
M$. In the large $M$ limit $U(1)_-$ decouples in the IR, and we are left with
the ${\cal N}=2$ supersymmetric $SU(M)\times SU(M)$ gauge theory with chiral
matter in $({\bf M},{\overline{\bf M}})$ and $({\overline {\bf M}},{\bf M})$.  

\section{Non-supersymmetric Theories}

{}So far we have been focusing on ${\cal N}=2$ and ${\cal N}=1$ supersymmetric
theories. However, in the large $N$ limit we can also discuss certain 
non-trivial non-supersymmetric cases as well. The large $N$ property is crucial
here. The reason is that in the cases where the orbifold group $\Gamma\not
\subset SU(3)$, we always have twisted NS-NS 
closed string sectors with tachyons. Their contributions to the
corresponding part of the annulus amplitude (\ref{NSNS}) is then exponentially 
divergent unless we require that 
\begin{equation}
 {\rm Tr}\left(\gamma_a\right)=0~,~~~g_a\not\in SU(3)~.
\end{equation}
Even if this condition is satisfied, we must take the 't Hooft limit - indeed,
otherwise it is unclear, for instance, how to deal with the diagrams with 
handles, which contain tachyonic divergences. In fact, the same applies to
some non-planar diagrams without handles, that is, diagrams where the 
external lines are attached to more than one boundaries (such diagrams are
subleading in the large $N$ limit).

{}To obtain well-defined non-supersymmetric non-conformal models, 
consider an orbifold group $\Gamma\subset Spin(6)$, which is not a 
subgroup of $SU(3)$. Let ${\widetilde \Gamma}$
be a non-trivial subgroup of $\Gamma$ such that ${\widetilde \Gamma}\subset
SU(2)$. We will allow the Chan-Paton matrices $\gamma_a$ corresponding to the
twists $g_a\in {\widetilde \Gamma}$ (which have $d_a=2$) not 
to be traceless, so that the corresponding
${\cal N}=2$ model is not conformal. However, we will require that
the other Chan-Paton matrices
$\gamma_a$ for the twists $g_a\not\in{\widetilde \Gamma}$ be traceless.
The resulting non-supersymmetric model is not conformal. However, in the planar
limit the perturbative gauge theory amplitudes are not renormalized 
beyond one loop (as usual, various running $U(1)$'s decouple in the IR in
this limit). The proof of this statement is completely parallel to that we
gave in subsection IIIA for ${\cal N}=1$ theories.

{}Let us consider a simple example of such a theory. Let $\Gamma\approx {\bf
Z}_2\otimes {\bf Z}_3$, where the action of the generators $R$ and $\theta$ of
the ${\bf Z}_2$ respectively ${\bf Z}_3$ subgroups on the complex coordinates
$z_\alpha$ on ${\cal M}={\bf C}^3/\Gamma$ is as follows: $R:z_1\rightarrow
z_1$, $R:z_{2,3}\rightarrow -z_{2,3}$, $\theta:z_1\rightarrow \omega z_1$, 
$\theta: z_{2,3}\rightarrow z_{2,3}$, where $\omega\equiv\exp(2\pi i/3)$. The
twisted Chan-Paton matrices are given by: $\gamma_R=I_{3N}$, $\gamma_\theta=
{\rm diag}(I_N,\omega I_N,\omega^{-1} I_N)$. Then the theory is a 
non-supersymmetric $U(N)\otimes U(N)\otimes U(N)$ gauge theory with
matter consisting of complex scalars in $({\bf N},{\overline{\bf N}},{\bf 1})$,
$({\bf 1},{\bf N},{\overline{\bf N}})$ and 
$({\overline{\bf N}},{\bf 1},{\bf N})$, as well as chiral fermions in the
above representations plus their complex conjugates.  

{}The gauge coupling renormalization in this model can be discussed in
complete parallel with the previous sections. Since the $U(1)$'s decouple in 
the IR in the large $N$ limit, we will ignore them in the 
following\footnote{As we saw in the previous sections, $U(1)$ runnings 
at one loop receive contributions from non-planar diagrams with the two 
external lines corresponding to the $U(1)$ gauge bosons attached to two
different boundaries. Note, however, that these contributions are due to
R-R exchanges (and the R-R sectors do not contain tachyons), so that 
tachyons do not contribute to this.}. 
Then the
relevant part of the classical action is given by (for simplicity
the matter field contributions are not shown, and we use the notation
$F^{\mu\nu}={\rm diag}(F^{\mu\nu}_1,F^{\mu\nu}_2,F^{\mu\nu}_3)$):
\begin{eqnarray}
 S=&&-
 \int_{\rm D3} \Big(a~{\rm Tr}\left[{\widehat F}^{\mu\nu}
 {\widehat F}_{\mu\nu}\right]
 +{1\over\sqrt{3}}
 \Big[b~\sigma~{\rm Tr}\left[\gamma_R~{\widehat F}^{\mu\nu}
 {\widehat F}_{\mu\nu}\right]+
 {\rm Tr}(\gamma_R)~{\widehat c}~\sigma\Big]\Big)\nonumber\\
 &&-\int_{{\rm D3}\times {\cal F}} 
 {1\over 2}\partial^\mu\sigma~\partial_\mu\sigma~,
\label{eq:actionu4}
\end{eqnarray}
where the couplings $a,b,{\widehat c}$ are the same as before, and the 
overall factor of $1/\sqrt{3}$ is due to the fact that the $R$ 
twisted field $\sigma$ now propagates in
${\bf R}^{1,3}\times {\cal F}$, where the fixed point set 
${\cal F}$ is an orbifold ${\bf R}^2/{\bf Z}_3$.

{}Integrating out the bulk fields, we obtain ($r=1,2,3$):
\begin{eqnarray}
 S^{\rm{\small qu}}_{\rm{\small brane}}=-\int {d^4 p\over (2\pi)^4}\sum_r
 {\widetilde a}_r(p^2)~
 {\rm Tr}\left[{\widehat F}^{\mu\nu}_r(p)~{\widehat F}_{r\mu\nu}(-p)\right]~, 
\end{eqnarray}
where the non-Abelian couplings are given by
\begin{equation}
 2{\widetilde a}_r={1\over g^2_{\rm {\small YM}}}+
 {\beta^{(r)}_0\over 16\pi^2}~\ln\left(
 {\Lambda^2\over p^2}\right)~.
\end{equation}
Here $\beta^{(r)}_0=-2N$ are the
one-loop $\beta$-function coefficients for the $SU(N)$ subgroups\footnote{Here
we should point out that, if the parent ${\cal N}=2$ theory has no matter
hypermultiplets, regardless of whether the final model has 
${\cal N}=0$ or ${\cal N}=1$ supersymmetry, in the above construction the
one-loop $\beta$-function coefficients for the non-Abelian subgroups are
always given by ${\widetilde \beta}_0/|\Gamma|$, where 
${\widetilde \beta}_0$ is the one-loop $\beta$-function coefficient of the
non-Abelian subgroup in the parent ${\cal N}=2$ theory.}.

\section{Comments on Non-perturbative Brane-Bulk Duality}

{}In the previous sections we discussed the brane-bulk duality in the
context of ${\cal N}=2$ as well as certain ${\cal N}=1$ and ${\cal N}=0$
large $N$ non-conformal gauge theories. In particular, we saw that 
in the planar limit perturbatively (on-shell) correlators in the 
corresponding ${\cal N}=0,1$ theories are the same as in the parent 
${\cal N}=2$ theories. Moreover, the one-loop
effective quantum action in such theories,
which in the large $N$ limit is not perturbatively renormalized beyond 
one-loop, can be computed by performing a classical computation in a higher
dimensional field theory. In this section we would like to discuss 
whether the non-perturbative corrections modify the brane-bulk duality picture
in such theories.

{}One of the key simplifying features here is the large $N$ limit. Thus, let
us consider the ${\cal N}=2$ supersymmetric $SU(N)$ gauge theory without 
matter. In this theory the low energy effective action can be described in 
terms of a prepotential ${\cal F}$, which perturbatively does not receive
corrections beyond one loop. The non-perturbative corrections come from 
instantons \cite{SW}:
\begin{equation}
 {\cal F}_{\rm{\small non-pert}}=\sum_{k=1}^\infty {\cal F}_k \Lambda_*^{2Nk}~,
\end{equation}
where $\Lambda_*$ is the dynamically generated scale of the theory:
\begin{equation}
 \Lambda_*=\mu\exp\left({8\pi^2\over g_{\rm{\small YM}}^2(\mu)\beta_0}\right)~.
\end{equation}
Here $g_{\rm{\small YM}}(\mu)$ is the Yang-Mills gauge coupling at some high 
scale $\mu$, and $\beta_0=-2N$ is the one-loop $\beta$-function coefficient.
Thus, the instanton corrections are weighted with
\begin{equation}
 \Lambda_*^{2Nk}=\mu^{2Nk}\exp\left(-{8\pi^2 N k\over \lambda(\mu)}\right)~,
\end{equation}  
where $\lambda(\mu)\equiv g_{\rm{\small YM}}^2(\mu)N$ is the effective 
't Hooft coupling. Note that these weights go to zero in the 't Hooft limit,
which implies that the low energy effective action is not renormalized beyond
one loop in the large $N$ limit. 

{}Next consider the ${\cal N}=1$ and ${\cal N}=0$ orbifold theories discussed
in subsection IIIA (as well as section VI) and section VII, respectively. 
Due to their underlying ${\cal N}=2$ structure, in these theories we might
hope that the non-perturbative corrections to the low energy effective 
action also vanish in the large $N$ limit. 
If so, then we have non-trivial statements
about infinitely many non-trivial ${\cal N}=0,1$ gauge theories, 
in particular, that in such theories the low energy effective action is not
renormalized beyond one loop in the planar limit. Checking this conjecture in 
the non-supersymmetric case is rather non-trivial, but in ${\cal N}=1$ cases
we can perform some partial checks. In particular, this conjecture implies that
in the large $N$ limit the superpotential should not receive non-perturbative
corrections, so that the classical superpotential should be exact as the 
superpotential does not receive any loop corrections in ${\cal N}=1$ 
supersymmetric theories. This statement can indeed be checked explicitly for 
such theories. Instead of being most general, here we will consider the 
simplest example of such an ${\cal N}=1$ theory
(other ${\cal N}=1$ cases can be discussed in a similar fashion). 
Thus, consider the example
discussed in section VI. In this example we have ${\cal N}=1$ supersymmetric
$SU(N)\otimes SU(N)$ gauge theory with chiral matter supermultiplets in 
$({\bf N},{\overline {\bf N}})$, 
and $({\overline{\bf N}},{\bf N})$. To simplify
the discussion, let us take the gauge coupling of the second $SU(N)$ factor
to be much smaller than that of the first one. Then the second $SU(N)$ can be
treated as the global symmetry group for the first $SU(N)$, and we have the
$SU(N)$ gauge theory with $N$ flavors of quarks $Q^i,{\overline Q}_{\bar
j}$, $i,{\bar j}=1,\dots,N$, where $Q^i$ and ${\overline Q}_{\bar
j}$ transform in the fundamental ${\bf N}$ respectively anti-fundamental
${\overline {\bf N}}$ of the gauge group $SU(N)$. The low energy dynamics is
described in terms of the gauge invariant degrees of freedom given by the
mesons $M^i_{\bar j}$ and baryons $B,{\overline B}$ \cite{seiberg}:
\begin{eqnarray}
 &&M^i_{\bar j}\equiv Q^i{\overline Q}_{\bar j}~,\\
 &&B\equiv \epsilon_{i_1\dots i_N}Q^{i_1}\cdots Q^{i_N}~,\\
 &&{\overline B}\equiv \epsilon^{\bar{j}_1\dots \bar{j}_N}
 {\overline Q}_{\bar{j}_1}\cdots {\overline Q}_{\bar{j}_N}~.
\end{eqnarray}
The classical moduli space in this theory receives quantum corrections, which 
can be accounted for via the following superpotential 
(here $A$ is a Lagrange multiplier related to the ``glueball'' field
via $A\Lambda_*^{2N}=W^\alpha W_\alpha$) \cite{seiberg}:
\begin{equation}
 {\cal W}_{\rm{\small{non-pert}}}=A\left(\det(M)-B{\overline B}-
 \Lambda_*^{2N}\right)~,
\end{equation}
where $\Lambda_*$ is the dynamically generated scale of the theory, which is
given by:
\begin{equation}
 \Lambda_*=\mu\exp\left(-{4\pi^2\over \lambda(\mu)}\right)~.
\end{equation}
Once again, $\Lambda_*^{2N}$ goes to zero in the large $N$ limit, so that the
classical constraint 
\begin{equation}
 \det(M)-B{\overline B}=0
\end{equation}
is unmodified in this limit. Thus, in this theory the classical
superpotential, which vanishes, is indeed exact in the large $N$ 
limit\footnote{On the other hand, it is not completely clear how to check
whether there are no non-perturbative corrections to the K{\"a}hler 
potential.}.

{}The above discussion suggests that the brane-bulk duality discussed in the
previous sections in the context of the aforementioned gauge theories might
hold even non-perturbatively, so that the corresponding low energy effective 
quantum action is not renormalized beyond one loop in the large $N$ limit. 

\section{Concluding Remarks}

{}We would like to end our discussion with a few concluding remarks. First,
one natural generalization we can consider is to extend the above discussion
to the cases containing $SO/Sp$ gauge 
groups. This can be done via orientifolding,
that is, by including orientifold planes in the setup of section II in the
spirit of \cite{z2}. In the ${\cal N}=2$ cases we expect no subtleties, but
in the ${\cal N}=1$ cases with twists with $d_a=0$ some caution is needed 
\cite{z2} due to the subtleties discussed in \cite{KaSh,KST}.

{}Another point we would like to comment on is the following. Recently, 
in the brane world context 
\cite{early,BK,pol,witt,lyk,shif,TeV,dien,3gen,anto,ST,BW,Gog,RS,DGP,DG,IK},
it was pointed out in \cite{DGP,DG} that the Einstein-Hilbert term is 
generically expected to be induced via loop corrections 
on a brane as long as the brane world-volume
theory is not conformal. Subsequently, it was argued in \cite{IK} that this 
effect should arise in the context of non-conformal gauge theories 
from D3-brane, in particular, this is expected to be the case in theories 
discussed in this paper. It would be interesting to see whether the 
brane-bulk duality can be used for simplifying computation of such 
corrections on the D3-branes.

\acknowledgments

{}We would like to thank Gregory Gabadadze, Rajesh Gopakumar, 
Martin Ro{\v c}ek, Tom Taylor, Henry Tye, Cumrun Vafa and Slava Zhukov 
for valuable discussions. This work was supported in part by the
National Science Foundation. Z.K. would also like to thank Albert and Ribena
Yu for financial support.

\appendix

\section{Brane-Bulk Couplings}

{}In section V we used various brane-bulk couplings in the ${\bf Z}_2$
orbifold examples. These couplings can be computed
within the boundary state formalism, where one computes the couplings of the 
boundary states to the twisted closed string states in the presence of 
non-trivial D-brane field backgrounds \cite{Nappi,BF,GG,DiVec}. Here,
however, since we only need the couplings relevant for the
one-loop corrections in the gauge theory language, we will take a shortcut and
deduce these couplings using the annulus amplitude in the presence of
D-brane field backgrounds. 

{}The annulus amplitude in the light-cone gauge is given by (here $R$ is the
generator of the ${\bf Z}_2$ orbifold twist):
\begin{equation}
 {\cal C}=\int_0^\infty {dt\over t} {\rm Tr}
 \left[{{1+R}\over 2}~{{1-(-1)^F}\over 2}~e^{-2\pi t L_0}\right]~.
\end{equation}
Here we assume that we have a non-trivial {\em constant} background gauge field
along the brane. The annulus amplitude ${\cal C}$ then is almost the same as
in the case without the background field, which we discussed in section II,
with the difference that some of the open string oscillator modings are 
modified \cite{Nappi}. 
For our purposes here it will suffice to consider the background
field of the form ($F_{\mu\nu}$, $\mu,\nu=0,1,2,3$, is the D3-brane 
gauge field strength):
\begin{equation}\label{eq:choiceF}
 F_{23}=BQ~,
\end{equation}
where $Q$ is a Cartan generator of the $U(N)$ gauge group, which we will take 
to be Hermitian. Let $q_\alpha$,
$\alpha=1,\dots,N$, be the eigenvalues of $Q$. It is convenient to 
introduce complex combinations of
the world-sheet bosonic and fermionic degrees of freedom corresponding to the
directions $\mu=2,3$ (so that instead of two real world-sheet bosons and 
two real world-sheet fermions corresponding to these directions we have one
complex world-sheet boson and one complex world-sheet fermion). Then the 
modings of these complex world-sheet degrees of freedom are modified as 
follows (modings of other world-sheet degrees of freedom are unchanged):
\begin{equation}
 r\rightarrow r+\Delta_{\alpha\beta}~,
\end{equation}
where $\Delta_{\alpha\beta}$ is given by:
\begin{equation}\label{Delta}
 \Delta_{\alpha\beta}={1\over \pi}\left[\arctan\left(2\pi\alpha^\prime 
 B q_\alpha\right) -
 \arctan\left(2\pi\alpha^\prime B q_\beta\right)\right]~.
\label{eq:epsilon}
\end{equation}
Here the indices $\alpha,\beta$ label the two ends of the open string.

{}The cylinder amplitude is given by \cite{BF}:
\begin{equation}
 {\cal C}=\int_0^\infty {dt\over t}~{1\over 8\pi^2\alpha^\prime t} 
 \sum_{\alpha,\beta} {(q_\alpha-q_\beta) B\over 2 \pi} \left[U_{\alpha\beta}+
 T_{\alpha\beta}\right]
 e^{-{t \over 2\pi\alpha^\prime}~X_{\alpha\beta}^2}~,
\end{equation}
where $X^2_{\alpha\beta}$ is square of the distance (in the two real extra 
directions untouched by the orbifold) between the D3-branes 
labeled by $\alpha$ and $\beta$. The factor of $8\pi^2\alpha^\prime t$ in the
denominator comes from the bosonic zero modes in the directions $\mu=0,1$.
The factor $(q_\alpha-q_\beta) B/2 \pi$ can be understood from the 
requirement that in the $B\rightarrow 0$ limit we must reproduce the 
corresponding answer (see below). 
In particular, the untwisted character $U_{\alpha\beta}$ and the twisted
character $T_{\alpha\beta}$ are given by\footnote{Here we should note that
there is some freedom in choosing the phases multiplying the terms containing
$Z^{-1/2}_{-1/2}$. Thus, for instance, we could have chosen the phase of 
the last term in the square brackets in (\ref{U}) as $-1$ instead of $+1$.
Note, however, that these phases do not affect our results here as the terms
they multiply are vanishing ($Z^{-1/2}_{-1/2}=0$).} 
(for simplicity the indices $\alpha,
\beta$ are not shown):
\begin{eqnarray}\label{U}
 U=&&{1\over 4 \eta^6(q)~
 Z_{-1/2}^{-1/2+\Delta}}\left[Z_0^\Delta ~[Z_0^0]^3 - Z_{-1/2}^\Delta ~
 [Z_{-1/2}^0]^3 \right.\nonumber\\
 &&\left.- Z_0^{-1/2+\Delta} ~[Z_0^{-1/2}]^3+
 Z_{-1/2}^{-1/2+\Delta} ~[Z_{-1/2}^{-1/2}]^3\right]~,\\
 \label{T}
 T=&&{1 \over 
 \eta^2(q) ~Z_{-1/2}^{-1/2+\Delta} ~[Z_0^{-1/2}]^2}\left[Z_0^\Delta ~Z_0^0 ~
 [Z_{-1/2}^0]^2 - Z_{-1/2}^\Delta ~Z_{-1/2}^0 ~[Z_{0}^0]^2\right.\nonumber\\
 &&\left.-Z_0^{-1/2+\Delta} ~Z_{-1/2}^0 ~[Z_{-1/2}^{-1/2}]^2+
 Z_{-1/2}^{-1/2+\Delta} ~Z_{-1/2}^{-1/2} ~[Z_{-1/2}^0]^2\right]~.
\end{eqnarray}
Here the characters $Z^v_u$ are the usual complex fermion characters
($-1/2\leq v <1/2$, $q\equiv\exp(-2\pi t)$):
\begin{equation}
 Z^v_u\equiv q^{{1\over 2}v^2-{1\over 24}} \prod_{m=1}^\infty
 \left(1+q^{m+v-{1\over 2}} ~e^{-2\pi i u}\right)~
 \left(1+q^{m-v-{1\over 2}} ~e^{2\pi i u}\right)~.
\end{equation}
In particular, note that the character $Z_{-1/2}^{-1/2+\Delta}$ in the 
denominators in (\ref{U}) and (\ref{T}) in the $B\rightarrow 0$ limit becomes:
\begin{equation}
 Z_{-1/2}^{-1/2+\Delta}\rightarrow 2\pi t \Delta~{1\over\eta^2(q)}~.
\end{equation} 
Combining this with the aforementioned factor $(q_\alpha-q_\beta) B/2 \pi$
gives the following contribution
\begin{equation}
 {(q_\alpha-q_\beta) B\over 4 \pi^2 \Delta_{\alpha\beta}~ t}~{1\over\eta^2(q)}
 \rightarrow {1\over 8\pi^2\alpha^\prime t}~{1\over\eta^2(q)}~,
\end{equation}
which precisely corresponds to the bosonic zero modes plus oscillators in the
$\mu=2,3$ directions in the absence of the gauge field background.

{}Here we would like to extract couplings of the massless closed string states
to the D-branes. This can be done by extracting the leading behavior of the
annulus amplitude for $t\rightarrow 0$. Using modular transformation 
properties of the above characters, we obtain:
\begin{eqnarray}
 &&U\sim {t^3\over \sin(\pi\Delta)} \left\{\left[2-\sin^2(\pi\Delta)
 \right]\Big|_{{\rm {\small NS-NS}}} - 
 2\cos(\pi\Delta)\Big|_{{\rm{\small R-R}}}\right\}~,\\
 &&T\sim {2t\over \sin(\pi\Delta)} \left\{1\Big|_{{\rm {\small NS-NS}}} - 
 \cos(\pi\Delta)\Big|_{{\rm{\small R-R}}}\right\}~,
\end{eqnarray}
where individual contributions due to the NS-NS and R-R exchanges are shown.

{}Here we are interested in the brane-bulk couplings involving at most 
quadratic terms in the gauge field strength $F_{\mu\nu}$. Then the relevant
terms in the small $B$ limit are given by:
\begin{eqnarray}
 &&U\sim {2t^3\over \pi\Delta} \left[1-{1\over 3}(\pi\Delta)^2\right]
 \left\{1\Big|_{{\rm {\small NS-NS}}} - 
 1\Big|_{{\rm{\small R-R}}}\right\}~,\\
 &&T\sim {t\over \pi\Delta} \left[1-{1\over 3}(\pi\Delta)^2\right]
 \left\{\left[2+(\pi\Delta)^2\right]\Big|_{{\rm {\small NS-NS}}} - 
 2\Big|_{{\rm{\small R-R}}}\right\}~.
\end{eqnarray}
Using the fact, which follows from (\ref{Delta}), that to the relevant order
\begin{equation}
 {1\over \pi\Delta_{\alpha\beta}} \left[1-{1\over 3}
 (\pi\Delta_{\alpha\beta})^2\right]={{1+(2\pi\alpha^\prime B)^2 
 q_\alpha q_\beta}\over (2\pi\alpha^\prime B) (q_\alpha-q_\beta)}~,
\end{equation} 
we obtain the following massless untwisted respectively massless twisted 
closed string contributions into the annulus amplitude
\begin{eqnarray}
 {\widetilde {\cal C}}_U=&&{1\over(4\pi^2\alpha^\prime)^2}
 \left\{1\Big|_{{\rm {\small NS-NS}}} - 
 1\Big|_{{\rm{\small R-R}}}\right\}\sum_{\alpha,\beta}
 \left[1+(2\pi\alpha^\prime B)^2 q_\alpha q_\beta\right]
 \int_0^\infty dt~t~e^{-{t\over 2\pi\alpha^\prime}~X_{\alpha\beta}^2}~,\\
 \label{T1}
 {\widetilde {\cal C}}_T=&&{1\over2(4\pi^2\alpha^\prime)^2}
 \sum_{\alpha,\beta}
 \left\{\left[2+
 (2\pi\alpha^\prime B)^2 \left(q_\alpha^2 +q_\beta^2\right)\right]
 \Big|_{{\rm {\small NS-NS}}}-
 \left[2+2(2\pi\alpha^\prime B)^2 q_\alpha q_\beta\right]
 \Big|_{{\rm{\small R-R}}}\right\}\nonumber\\
 &&\int_0^\infty 
 {dt\over t}~e^{-{t\over 2\pi\alpha^\prime}~X_{\alpha\beta}^2}~.
\end{eqnarray}
Note that the integrals in these expressions are related to the corresponding
Euclidean propagators $\Delta_6(X_{\alpha\beta}^2)$ and 
$\Delta_2(X_{\alpha\beta}^2)$:
\begin{equation}
 \Delta_d(y^2)\equiv \int {d^d p\over (2\pi)^d}~{e^{ip\cdot y}\over p^2}=
 {1\over 4\pi^{d/2}}\int_0^\infty~ds~s^{d-4\over 2} ~e^{-s y^2}~.
\end{equation}
here $y^2\equiv y^i y^i$, $i=1,\dots,d$, and
$p^i$ are the momenta corresponding to the coordinates $y^i$.

{}As we discussed in section V, the relevant part of the classical action is
given by (note that for the field of the form (\ref{eq:choiceF}) 
terms containing $F\wedge F$ are vanishing):
\begin{eqnarray}
 S=&&-
 \int_{\rm D3} \Big(a~{\rm Tr}\left[F^{\mu\nu}F_{\mu\nu}\right]+
 b~\sigma~{\rm Tr}\left[\gamma_R~F^{\mu\nu}F_{\mu\nu}\right]+
 {\rm Tr}(\gamma_R)~{\widehat c}~\sigma+
 d~\epsilon^{\mu\nu\sigma\rho}~C_{\mu\nu}~
 {\rm Tr}\left[\gamma_R~F_{\sigma\rho}\right]\Big)\nonumber\\
 &&-\int_{{\rm D3}\times {\bf R}^2} 
 \left({1\over 2}\partial^\mu\sigma~\partial_\mu\sigma+{1\over 12}
 H^{\mu\nu\sigma} H_{\mu\nu\sigma}\right)~.
\label{eq:actionuA}
\end{eqnarray}
Here $\sigma$ is a twisted NS-NS scalar, while $C_{\mu\nu}$ is a twisted
two-form (whose field strength is $H_{\mu\nu\sigma}$). The twisted Chan-Paton
matrix $\gamma_R=I_N$. Note, however, that for generic values of 
$X_{\alpha\beta}$ the $U(N)$ gauge group is broken down to $U(1)^N$. So
the D3-branes are not necessarily coincident in the two transverse dimensions
untouched by the orbifold action.

{}From the tree-level action (\ref{eq:actionuA}) we obtain the following
twisted massless contributions quadratic in the gauge field strength
(here we are using $F^{\mu\nu}F_{\mu\nu}=2B^2 Q^2$):
\begin{equation}
 {\widetilde {\cal C}}_T^\prime=2b{\widehat c}\sum_{\alpha,\beta}
 2B^2 \left(q_\alpha^2+q_\beta^2\right)\Delta_2(X_{\alpha\beta}^2)\Big|_\sigma
 -8d^2 \sum_{\alpha,\beta} 2B^2 q_\alpha q_\beta
 \Delta_2(X_{\alpha\beta}^2)\Big|_{C_{\mu\nu}}~.
\end{equation}  
Note that this expression contains an overall factor of 2 from exchanging the
two ends of the string as we are discussing an {\em oriented} string theory.
Comparing this expression with (\ref{T1}), we obtain:
\begin{equation}
 b{\widehat c}=2d^2={1\over 8\pi}~. 
\end{equation}
The coupling ${\widehat c}$ can be determined in a similar fashion.

\newpage
\begin{figure}[t]
\epsfxsize=15 cm
\epsfbox{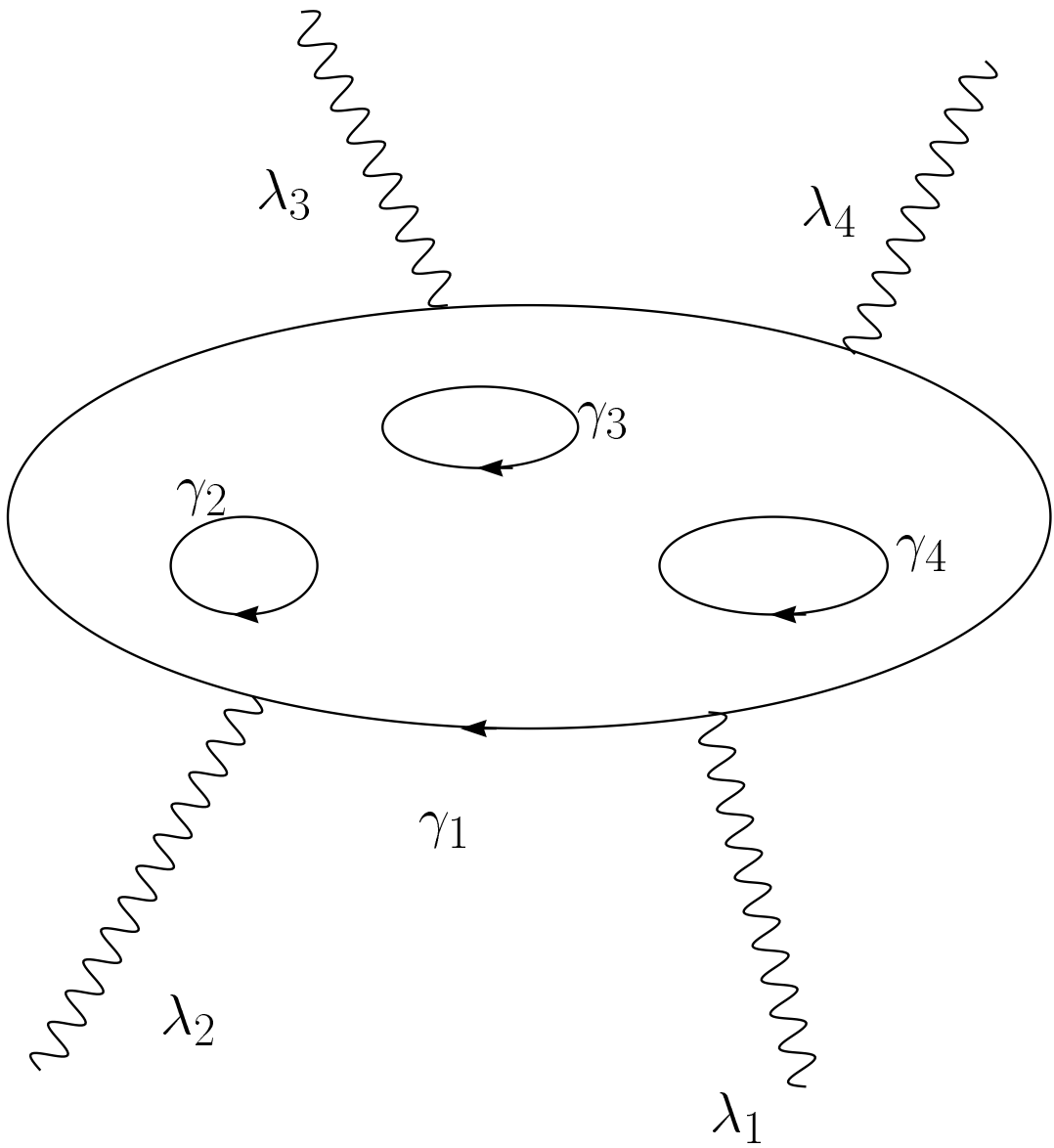}
\caption{A planar diagram.}
\end{figure}

\end{document}